\documentclass[twosided]{emulateapj}
\usepackage{ifthen}
\usepackage{amsmath,amssymb}
\usepackage{graphicx,mathptm}
\usepackage{times}
\usepackage{natbib}
\usepackage[usenames]{color}
\usepackage{ifpdf}
\usepackage{thumbpdf}
\usepackage{epstopdf}
\usepackage{pdflscape}
\usepackage{thumbpdf}

\newcommand{\beq}{
\begin{equation}
}
\newcommand{\eeq}{
\end{equation}
}
\newcommand{\beqa}{
\begin{eqnarray}
}
\newcommand{\eeqa}{
\end{eqnarray}
}

\newcounter{emulateapj}
\makeatletter
\@ifclassloaded{emulateapj}
{
\setcounter{emulateapj}{1}
}
{
\setcounter{emulateapj}{0}
}
\makeatother

\newcommand{\figdir}{./}

\newcommand{\msun}     {\ensuremath{{\rm M}_{\scriptscriptstyle \odot}}}

\newcommand{\ergs}     {\ensuremath{~\mathrm{erg\,s^{-1}}}}

\newcommand{\Mpc}      {\ensuremath{~\mathrm{Mpc}}}
\newcommand{\msigma}   {\ensuremath{M}{--}\ensuremath{\sigma}}
\newcommand{\ml}       {\ensuremath{M}{--}\ensuremath{L}}
\newcommand{\mbh}      {\ensuremath{M_{\mathrm{BH}}}}

\newcommand{\lx}      {\ensuremath{L_X}}
\newcommand{\lr}      {\ensuremath{L_R}}
\newcommand{\ledd}      {\ensuremath{L_{\mathrm{Edd}}}}
\newcommand{\fedd}      {\ensuremath{f_{\mathrm{Edd}}}}

\newcommand{\lrscale}  {\ensuremath{L_{R,38}}}
\newcommand{\lxscale}  {\ensuremath{L_{X,40}}}
\newcommand{\mbhscale}      {\ensuremath{M_{\mathrm{BH},8}}}

\ifthenelse{\value{emulateapj} = 1}
{}
{
\newenvironment{deluxetable*} {\begin{deluxetable}}{\end{deluxetable}}
}

\def\spose#1{\hbox to 0pt{#1\hss}}
\newcommand{\lta}{\mathrel{\spose{\lower 3pt\hbox{$\mathchar"218$}}
      \raise 2.0pt\hbox{$\mathchar"13C$}}}
\newcommand{\gta}{\mathrel{\spose{\lower 3pt\hbox{$\mathchar"218$}}
      \raise 2.0pt\hbox{$\mathchar"13E$}}}
\def\simlt{\mathrel{\rlap{\lower 3pt\hbox{$\sim$}}\raise 2.0pt\hbox{$<$}}}
\def\simgt{\mathrel{\rlap{\lower 3pt\hbox{$\sim$}} \raise 2.0pt\hbox{$>$}}}

\definecolor{KayhanCiteColor}{rgb}{0,0.0,0.0}
\definecolor{KayhanURLColor}{rgb}{0,0.08,0.35}
\definecolor{KayhanLinkColor}{rgb}{0,0.0,0.0}
\definecolor{KayhanPageColor}{rgb}{0,0.0,0.0}

\shorttitle{The Fundamental Plane with Dynamical Masses} 
\shortauthors{G\"{u}ltekin et al.}

\makeatletter \ifx\undefined\hyperref \usepackage[pdftitle={The
Fundamental Plane of Accretion onto Black Holes with Dynamical
Masses},pdfauthor={Kayhan
Gultekin},pdfsubject={Astrophysics},pdfkeywords={black hole physics
--- galaxies: general --- galaxies:
nuclei},letterpaper,linktocpage,colorlinks=true,linkcolor={KayhanLinkColor},pagecolor={KayhanPageColor},urlcolor={KayhanURLColor},citecolor={KayhanCiteColor},breaklinks=true]{hyperref}
\else\relax\fi \makeatother
\ifthenelse{\value{emulateapj} = 1}
{
\usepackage[letterpaper,paperwidth=8.5in,paperheight=11.0in,width=6.5in,height=9.0in,inner=1.0in,outer=0.375in,top=1.5in,bottom=-0.25in]{geometry}
}
{}
\begin{document}

\label{firstpage}

\title{The Fundamental Plane of Accretion onto Black Holes with Dynamical Masses}

\author{Kayhan G\"{u}ltekin\altaffilmark{1}}
\author{Edward M. Cackett\altaffilmark{1,4}}
\author{Jon M. Miller\altaffilmark{1}}
\author{Tiziana Di Matteo\altaffilmark{2}}
\author{Sera Markoff\altaffilmark{3}}
\author{Douglas O. Richstone\altaffilmark{1}}
\affil{\altaffilmark{1}Department of Astronomy, University of Michigan, Ann Arbor, MI, 48109.  Send correspondence to \href{mailto:kayhan@umich.edu}{kayhan@umich.edu}.}
\affil{\altaffilmark{2}McWilliams Center for Cosmology, Physics Department, Carnegie Mellon University, Pittsburgh, PA, 15213.}
\affil{\altaffilmark{3}A. Pannekoek, University of Amsterdam, 1090GE Amsterdam, NL.}
\altaffiltext{4}{Chandra Fellow}

\begin{abstract}

Black hole accretion and jet production are areas of intensive study in 
astrophysics.  Recent work has found a relation between radio luminosity, 
X-ray luminosity, and black hole mass.  With the assumption that radio and 
X-ray luminosity are suitable proxies for jet power and accretion power, 
respectively, a broad fundamental connection between accretion and jet 
production is implied.  In an effort to refine these links and enhance their 
power, we have explored the above relations exclusively among black holes with 
direct,
dynamical mass-measurements.  This approach not only eliminates
systematic errors incurred through the use of secondary mass
measurements, but also effectively restricts the range of distances
considered to a volume-limited sample.  Further, we have exclusively
used archival data from the \emph{Chandra X-ray Observatory} to best
isolate nuclear sources.  We find $\log{\lr} = \left(4.80 \pm
0.24\right) + \left(0.78 \pm 0.27\right) \log{\mbh} + \left(0.67 \pm
0.12\right) \log{\lx}$, in broad agreement with prior efforts.
Owing to the nature of our sample, the plane can be turned into an
effective mass predictor.  When the full sample is considered, masses
are predicted less accurately than with the well-known \msigma\
relation.  If obscured AGN are
excluded, the plane is potentially a better predictor than other
scaling measures.

\end{abstract}

\section{Introduction}
\setcounter{footnote}{5}
\label{intro}

Accretion onto black holes has many observable consequences, including
the production of relativistic jets.  The phenomenon of jet production
appears to be universal, as such jets are observed both in Active
Galactic Nuclei (AGN) and stellar-mass black hole systems as well as
in neutron stars, white dwarfs, and even young stellar objects.  For
black hole sources, the length scales and relevant timescales of jets
appear to approximately scale with mass over 8 orders of magnitude,
giving rise to the possibility that jet production mechanisms scale
with mass, similar to the way that accretion disk properties scale.
The mechanism by which jets are driven from black holes, however,
remains observationally elusive.  It remains one of the most
compelling and important problems in astrophysics, particularly in
high energy astrophysics.  The impact of relativistic jets on the
interstellar medium \citep{2005Natur.436..819G}, and large-scale
structure in clusters of galaxies
\citep{2006MNRAS.372...21A,2003MNRAS.344L..43F,2006ApJ...648..164M},
has become dramatically clear in the era of imaging and spectroscopy
with \emph{Chandra}.

Virtually all theories of jet production tie the jet to the accretion
disk directly or indirectly \citep[see, e.g.,][see also
\citealt{2009arXiv0905.3367V}]{1978PhyS...17..185L,1982MNRAS.199..883B}.
Thus, there is a broad expectation that jet properties might depend on
the mass accretion rate ($\dot{M}$) through the disk.  The black hole
spin parameter ($a \equiv cJ/GM^2$; $0 < a < 1$) may also be an
important factor if the black hole and accretion disk are linked
through magnetic fields \citep{1977MNRAS.179..433B}.  The spin is also
important for accretion disk jet-launching because the inner radius of
the accretion will decrease, thus increasing the launch velocity.
This idea may find some support in the dichotomy between radio-loud
and radio-quiet AGN \citep{2007ApJ...658..815S}.  The high flux of
stellar-mass black holes facilitates spin constraints with current
X-ray observatories; in those systems, the most relativistic jets
appear to be launched by black holes with high spin parameters
\citep{milleretal09}.

One means by which jet production can be examined is to explore
correlations between proxies for mass inflow and jet outflow.  In
stellar-mass black holes, it was found that radio emission and X-ray
emission are related by $\lr \propto \lx^{0.7}$ \citep{galloetal03}.
This correlation was quickly extended to also include super-massive
black holes in AGN, resulting in the discovery of a ``fundamental
plane'' of black hole activity \citep[][also see
\citealt{merlonietal06}]{merlonietal03,fkm04}.  The plane can be
described by $\log\lr = 7.33 + 0.60\log\lx + 0.78\log\mbh$, with a
scatter of $\sigma_R = 0.88\ \mathrm{dex}$ (where \lr\ is $\nu = 5\
\mathrm{GHz}$ nuclear radio luminosity in units of \ergs, \lx\ is $E =
2$--$10\ \mathrm{keV}$ nuclear X-ray luminosity in units of \ergs, and
\mbh\ is the black hole's mass in units of \msun;
\citealt{merlonietal03,fkm04}).  Several recent works have revisited
the original findings with slightly different focuses.
\citet{2006A&A...456..439K} found that sources emitting far under
their Eddington limits followed the relation more tightly.
\citet{2006ApJ...645..890W} found differences in the relationship for
radio-loud and radio-quiet AGNs.  \citet{2008ApJ...688..826L} used a
large sample of SDSS-identified broad-line AGNs to study a similar
relation at lower-frequency (1.4 GHz) radio luminosity and softer-band
(0.1--2.4 keV) X-ray luminosities.  \citet{2009arXiv0902.3704Y}
limited the sample to those sources with $\lx / \ledd < 10^{-6}$ based
on predictions that the correlation between radio and X-ray luminosity
steepens to $\lr \propto \lx^{1.23}$ at low accretion rates
\citep{2005ApJ...629..408Y}.

It is difficult to overstate the potential importance of the
fundamental plane; it suggests that black holes regulate their
radiative and mechanical luminosity in the same way at any given
accretion rate scaled to Eddington, $\dot{m} = \dot{M} /
\dot{M}_\mathrm{Edd}$.  In the context of models that assume jet
properties do scale simply with mass
\citep[e.g.,][]{1995A&A...293..665F,2003MNRAS.343L..59H}, the
fundamental plane can even be used to constrain the nature of the
accretion inflow.  At present, radiatively-inefficient inflow models
for X-ray emission, and models associating X-ray flux with synchrotron
emission near the base of a jet, are both consistent with the
fundamental plane.

To use the fundamental plane as a tool and a diagnostic instead of as
an empirical correlation, however, it must be sharpened.  Black hole
masses represent a significant source of uncertainty and scatter in
the fundamental plane \citep{merlonietal03,2006A&A...456..439K}.  In
this work, we have constructed a fundamental plane using only black
holes with masses that have been dynamically determined, the so-called
\msigma\ black holes \citep[see][]{gultekinetal09b}.  Unlike prior
treatments, our X-ray data is taken from a single observatory and
predominately from a single observing mode, and we have conducted our
own consistent analysis of the data.  We analyzed every archival {\it
Chandra} X-ray observation of black holes with a
dynamically-determined mass.  Radio data were taken from archival
observations reported in the literature.  By using a sample of black
holes with dynamical masses, we may probe the fundamental plane
without subjecting the analysis to the systematic errors inherent in
substituting scaling-relation-derived quantities for black hole
masses.

In Section~\ref{sample} we describe the sample of black holes used in
this work.  We detail our X-ray data reduction and spectral fits in
Section~\ref{xrayanal}.  Our fitting methods and results are presented
in Section~\ref{anal}.  We discuss our results in
Section~\ref{discussion} and summarize in Section~\ref{concl}.

\section{Sample of Black Holes}
\label{sample}
We get \mbh\ from the list of black hole masses compiled in
\citet{gultekinetal09b}, adopting the same distances as well.  This
sample of black hole masses includes measurements based on high
spatial resolution line-of-sight stellar velocity measurements
\citep[e.g.,][]{gultekinetal09a}, stellar proper motions in our Galaxy
\citep{ghezetal08,2008arXiv0810.4674G}, gas dynamical measurements
\citep[e.g.,][]{barthetal01}, and maser measurements
\citep[e.g.,][]{1995Natur.373..127M}.  It does not include
reverberation mapping measurements, which are direct measurements of
mass but are secondary in that they are normalized to the other
measurements via the \msigma\ relation
\citep[e.g.,][]{petersonetal04,onkenetal04}.  From those available
black hole masses we use only the measured black hole masses used in
their \msigma\ fits---not upper limits and not the ``omitted sample,''
which contains a list of masses with potential problems because (1)
masses were listed as tentative by the original study, (2) there was
no quantitative analysis of how well the original study's model fit
the data, or (3) the quantitative analysis of the goodness of fit was
poor.  We reduce available \emph{Chandra} data and present X-ray
luminosities for this collection of potentially problematic masses,
but we do not use them in our fits.  Thus, we use only the black hole
masses with the most reliable measurements.

One of the benefits of using this sample is that most of the distances
to the galaxies are less than $30\ \Mpc$.  This distance is close
enough that interestingly low X-ray and radio luminosities will still
be measurable.  So while this is not a true volume-limited sample, it
is insensitive to the potential biases arising from, e.g., a sample
limited by X-ray flux.  Unlike other samples, however, our sample may
be biased to very low nuclear luminosities.  The contamination from a
bright AGN typically causes problems in determining stellar
mass-to-light ratio at the center so that most galaxies selected for
dynamical measurement do not contain bright AGN.  Spiral galaxies,
which are less massive on average than early-type galaxies, may also
be underrepresented in this sample, and thus low-mass black holes may
also be underrepresented.

The radio data we use are 5 GHz peak power measurements from the
\citet{ho02} compilation of nuclear radio sources.  The data were
compiled to probe whether there was a correlation between \mbh\ and
\lr and thus are ideal for our purposes.  

\section{X-ray Analysis}
\label{xrayanal}

\subsection{X-ray Data Reduction}
\label{xrayreduce}
The high spatial resolution of \emph{Chandra} enables nuclear emission
to be isolated best compared to other X-ray observatories.  We used
\emph{Chandra} archival data to obtain accurate measurements or tight
upper limits of the flux between 2 and 10 keV for all galaxies in our
sample.  This energy range was chosen to probe accretion power rather
than total power including any contaminating diffuse emission and for
ease in comparison to previous fundamental plane work.

For each source we used a circular extraction region positioned at the
brightest point source that was consistent with the center of galaxy
determined by 2MASS images.  There are no nuclear point sources in
NGC~1399 and NGC~4261, which we handled slightly differently as
described below.  For extraction of background spectra, we typically
used an annular region with inner radius slightly larger than the
source region radius.  The outer radius was made large enough to
encompass a significant number of counts.  Point sources were excluded
from the background region.  When there were a large number of point
sources in the annular region surrounding the source, a different
region was used, usually an off-nuclear circle.  In these cases we
selected a region where the background looked to be similar to that
surrounding the source.

Two cases require special attention to contamination from non-nuclear
X-ray emission: NGC~0224 (M31) and NGC~4486 (M87).  NGC~0224 has two
bright point sources near the center of the galaxy, but neither is the
galaxy's central black hole, from which the emission is too dim to be
detected above the background to high significance \citep[$L_X \lta
10^{36}\,\ergs$ at at assumed distance of $D =
0.8\,\Mpc$][]{2005ApJ...632.1042G,2009arXiv0902.3847L}.  NGC~4486 is
well known for its prominent jet with several knots.  These knots are
apparent in the \emph{Chandra} images. In the high spatial resolution
\emph{Hubble Space Telescope} (\emph{HST}) images, one knot is very
close \citep[$0\farcs85$][]{2006ApJ...640..211H} to the central
engine.  As the knot has grown brighter in the optical by a factor of
$\sim100$ over the last $\sim10$ years, measurements of the core X-ray
flux become increasingly contaminated by the knot.  We chose the
archival \emph{Chandra} data set where the knot was most readily
distinguishable from the core.

For the galaxies NGC~1399 and NGC~4261, there is no discernible point
source at their nuclei, which are dominated in X-rays by hot gas.  For
these two sources, we attempt to measure a hypothetical point source
at the center.  We use a circular region at the center of the diffuse
X-ray emission for source extraction with an annular background
extraction region immediately adjacent.  For both these sources, X-ray
point source flux could not be inferred above the background, and they
are listed as upper limits in Table~\ref{t:data}.

Data reduction followed the standard pipeline, using the most recent
\emph{Chandra} data reduction software package (CIAO version 4.1.1)
and calibration databases (CALDB version 4.1.2).  Point-source spectra
were extracted using the CIAO tool psextract.  Because all
observations of interest were done with the Advanced CCD Imaging
Spectrometer (ACIS), we ran psextract with the mkacisrmf tool to
create the response matrix file (RMF) and with mkarf set for ACIS
ancillary response file (ARF) creation.

\subsection{X-ray Spectral Fitting}
\label{xrayfit}

We modeled the reduced spectra using XSPEC12
\citep{1996ASPC..101...17A}.  If binning the spectra in energy so that
each bin contained a minimum of 20 counts resulted in five or more
bins, we did so and used $\chi^2$ statistics; otherwise we did not bin
the data and used $C$-stat statistics \citep{1979ApJ...228..939C}.
Each spectrum was modeled with a photoabsorbed power-law model.  If
such a model did not adequately fit the spectrum for data sets that
were strong enough to support a more complicated model, we added
additional model components.  Galaxies that were identified as Seyfert
2 or transitional Seyferts in \citet{2006A&A...455..773V} were modeled
with a partially photoabsorbed power-law, representing intrinsic
absorption plus another photoabsorbed component, representing Galactic
absorption.  Galaxies with obvious diffuse hot gas towards their
nucleus were modeled with photoabsorbed Astrophysical Plasma Emission
Code \citep[APEC][]{2001ApJ...556L..91S} and power-law components.
Regardless of the continuum model, for spectra that showed an obvious
Fe K$\alpha$ line, we added a Gaussian for each line.  All spectra
were fit from $E = 0.5$ to $10~\mathrm{keV}$.

We considered a model successful if it yielded a reduced $\chi^2$ of
$\chi^2/\nu \le 2$ and if the spectrum between $E = 2$ and $10\ 
\mathrm{keV}$ was adequately described.  The total flux between $E =
2$ and $10\ \mathrm{keV}$, $F_{X,\mathrm{tot}}$, was determined from
the model and the 1$\sigma$ errors derived from covariance of the
model parameters.  We then calculated the unabsorbed flux arising from
just the power-law component between $E = 2$ and $10\ \mathrm{keV}$,
$F_{X}$.  That is, we de-absorbed the flux and removed contributions
from lines and other model components.  We assume the fractional error in
$F_{X,\mathrm{tot}}$ is the same as in $F_X$.

For sources that did not constrain the flux from the central point
source, we used the total count rate between $E = 0.5$ and $10
\mathrm{keV}$ to calculate the 3$\sigma$ (99.7\% confidence) upper
limit to $F_X$ with PIMMS assuming a power-law with index $\Gamma = 2$
and with Galactic absorption determined from the Leiden/Argentine/Bonn
survey of Galactic HI
\citep{2005A&A...440..775K,1997agnh.book.....H,2005A&A...440..767B}
using the HEASOFT ftool ``N$_{\rm H}$''.

\begin{figure*}[t]
\centering
\includegraphics[width=0.45\textwidth]{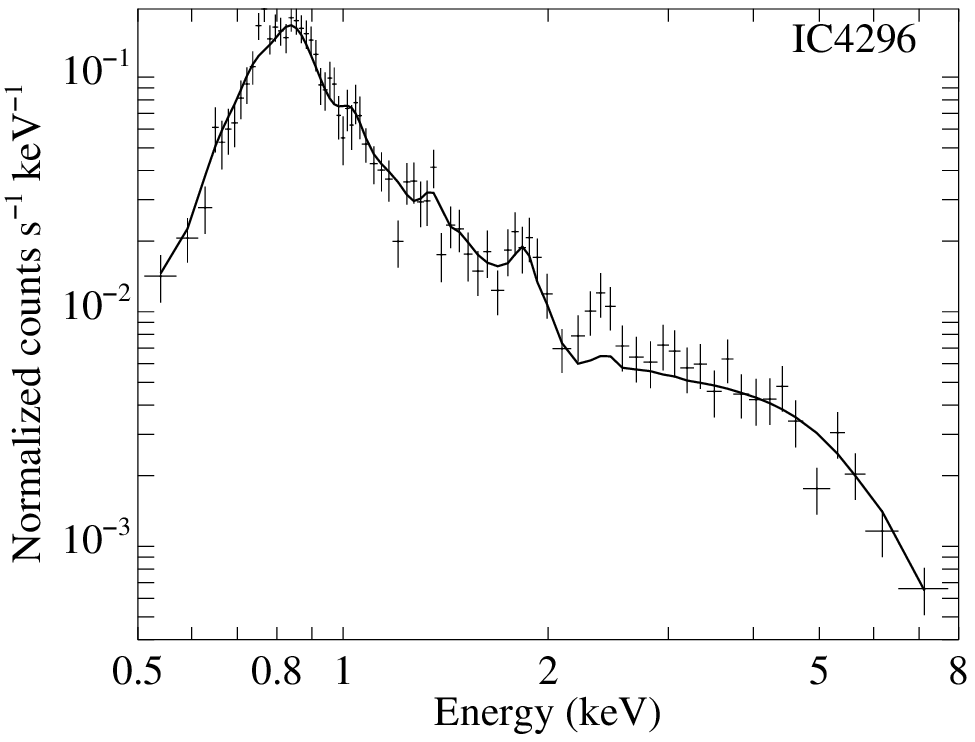}
\includegraphics[width=0.45\textwidth]{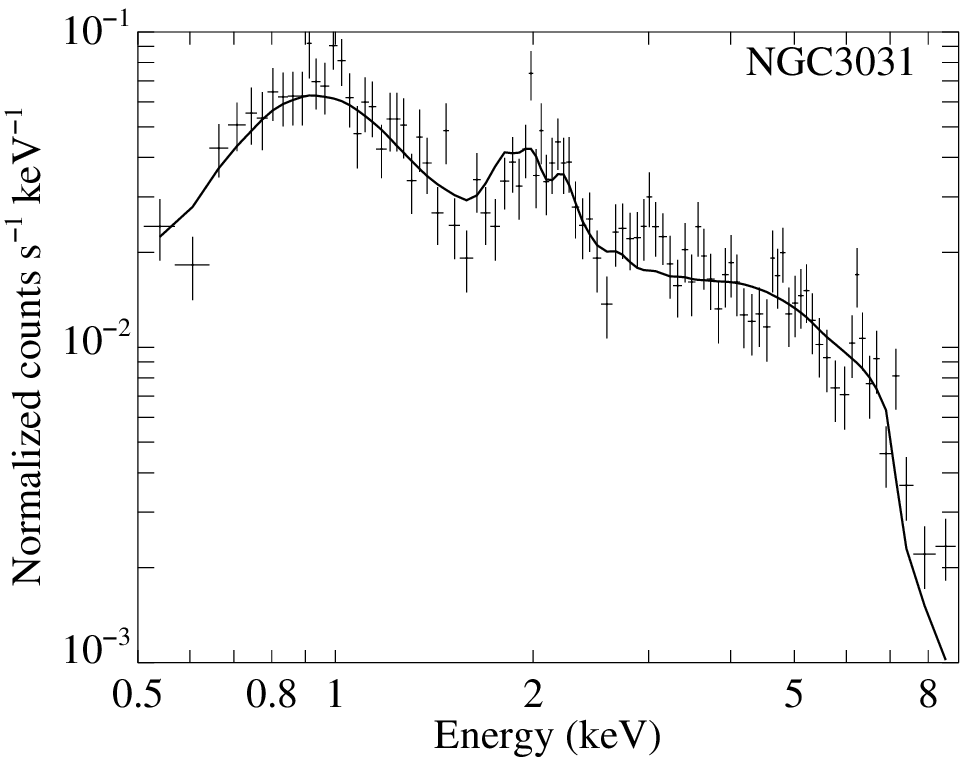}\\
\includegraphics[width=0.45\textwidth]{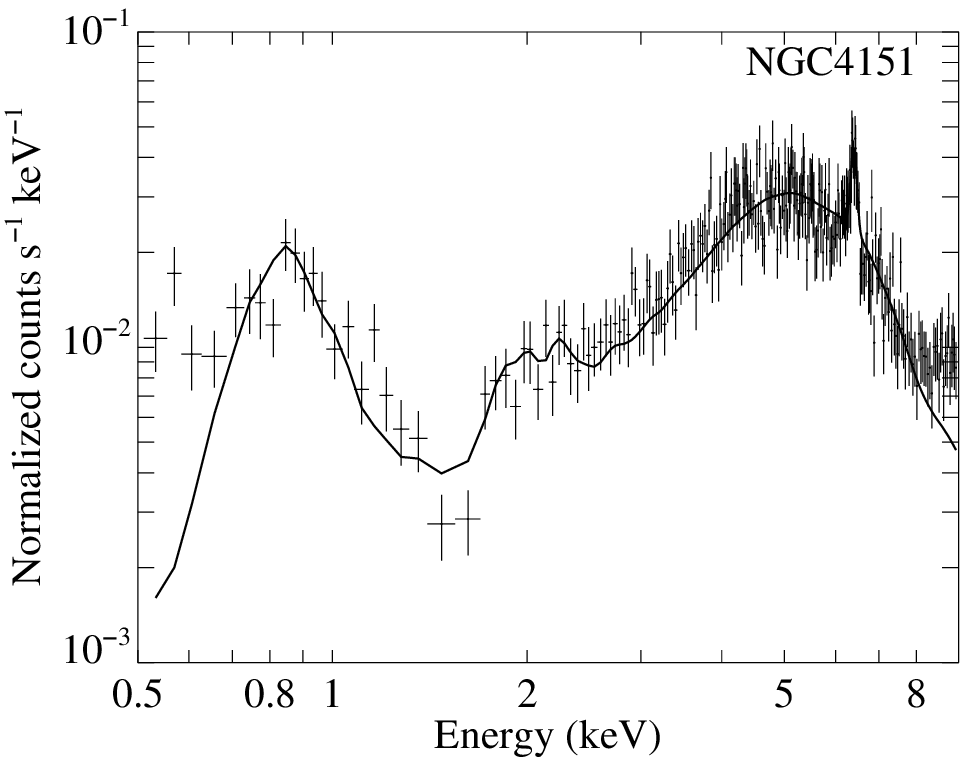}
\includegraphics[width=0.45\textwidth]{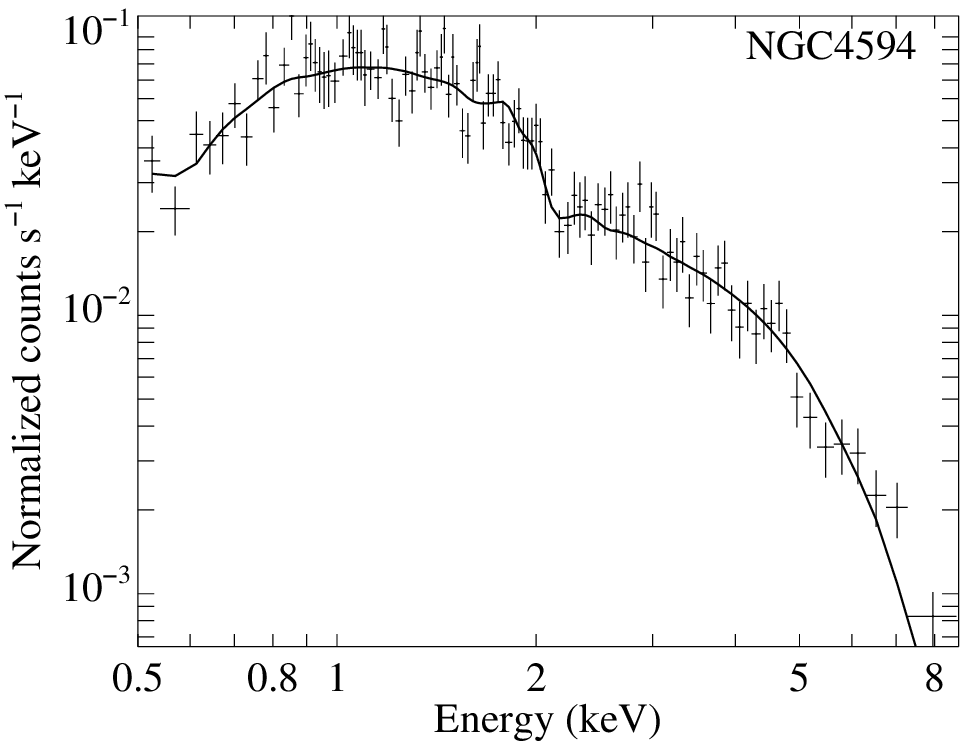}
\caption{Example \emph{Chandra} spectra with best-fit models.  The
models have been folded through the instrument response.  The
horizontal error bars show the binning used for the fits.  These four
galaxies were chosen to show a variety of different models used to fit
the data.  All spectra included Galactic absorption and a power-law
component.  NGC~3031, NGC~4151, and NGC~4594 included intrinsic
absorption; IC~4296 and NGC~4151 included an APEC model; and NGC~4151
included a Gaussian component to model the Fe line.}
\label{f:xfits}
\end{figure*}

Because we are ultimately interested in an accurate measurement of
$F_X$, it is more important that our models characterize the spectrum
well over the 2 to 10 keV band than it is to reproduce the underlying
physics.  We tested this approach by fitting several different models
to the same spectrum and recovered consistent values for $F_X$.  The
results of fits are displayed in Table~\ref{t:xray}, and we show four
example spectra with models in Figure~\ref{f:xfits}.

For many galaxies, multiple \emph{Chandra} observations were available
in the archive.  We reduced and analyzed the available data and
censored the resulting data by (1) choosing those that yielded flux
detections as opposed to upper limits, (2) choosing those with smaller
values of $\chi^2/\nu$, (3) preferring higher precision measurements
over lower precision, and (4) observed more closely in time with the
available radio data since variable sources will have \lr\ and \lx\
change in concert on the fundamental plane \citep[see][]{merlonietal06}.

\begin{figure}[th]
\includegraphics[width=\columnwidth]{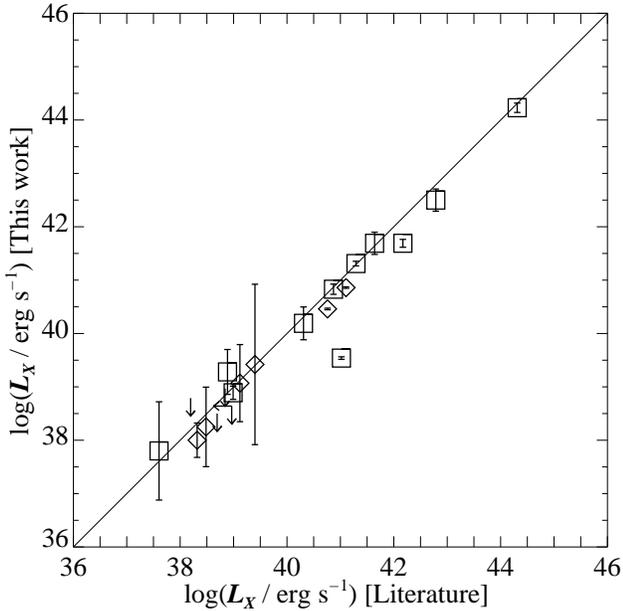}
\caption{Comparison of results of X-ray analysis in this work to
results from the literature.  All values have been scaled to our
adopted distances.  Squares indicate straight-forward comparisons.
Diamonds indicate that we have converted the literature result to an
unabsorbed 2--10 keV luminosity using the published spectral fit and
absorption.}
\label{f:complum}
\end{figure}

We compare our results with results from the literature for the same
data sets in Figure~\ref{f:complum}.  The literature values were
scaled to our assumed distances and, in some cases, converted to the
2-10 keV band with PIMMS and the published spectral fits.  The
comparison reveals good agreement with no particular bias with
exception of a single outlier, NGC~1068.  We expand on NGC~1068 and
Compton-thick sources in general below.

\begin{figure}[t]
\centering
\includegraphics[width=1.0\columnwidth]{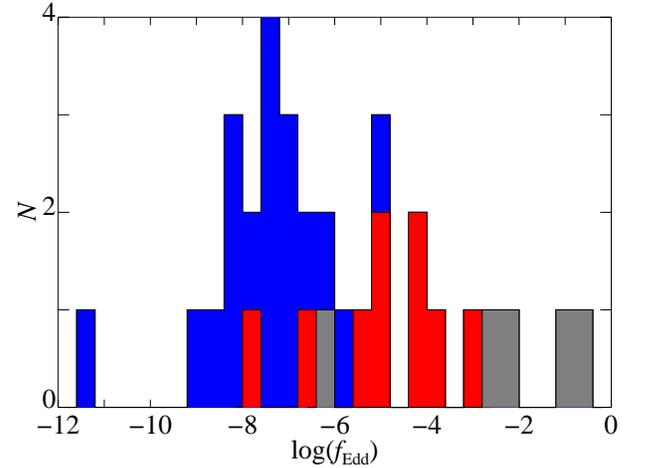}
\caption{Histogram of Eddington fractions defined as $\fedd = \lx /
\ledd$.  The contribution to the histogram from Seyfert galaxies is
colored red, from other SMBH sources is colored blue, and from
stellar-mass sources considered in section~\ref{stellarmass} is
colored gray.  The galaxy with the smallest $\fedd$ is Sgr~A*.  A wide
range of values are present in the sample even if most are found
between $\fedd = 10^{-9}$ and $10^{-6}$.  As expected, galaxies
classified as Seyferts are, on average, emitting at a higher fraction
of Eddington than other sources, and the stellar-mass sources are
emitting at a higher fraction still.}
\label{f:fedd}
\end{figure}

For the Milky Way (Sgr~A*) we used the literature result from
\citet{2001Natur.413...45B} during quiescence.  The data we use are
displayed in Table~\ref{t:data} along with other galaxies with
dynamically measured black holes without measurements of \lx, \lr, or
either.  A summary of the X-ray analysis may be gleaned from
Figure~\ref{f:fedd}, which shows a histogram of values of Eddington
fractions $\fedd = \lx / \ledd$ for all objects that resulted in an
X-ray measurement.  The distribution shows that while most are
accreting at a small fraction of Eddington, there are still a wide
range of values encompassed in the sample.

\section{Analysis}
\label{anal}
\subsection{Fitting Method}
\label{fitmeth}
For our measurement of the relation between \mbh, \lr\ and \lx, we
considered the form
\beq
\log{\lrscale} = R_0 + \xi_m \log{\mbhscale} + \xi_x \log{\lxscale},
\label{e:fitfunc}
\eeq
where we have normalized to $\lr = 10^{38} \ergs \lrscale$, $\mbh =
10^{8} \msun \mbhscale$, and $\lx = 10^{40} \lxscale$ in order to
minimize intercept errors.
To find the multi-parameter relation, we minimized the following statistic
\beq
\tilde{\chi}^2 = \sum_i{\frac{\left(R_i - R_0 - \xi_m \mu_i - \xi_x X_i\right)^2}
{\sigma_{r,i}^2 + \xi_m^2 \sigma_{m,i}^2 + \xi_x^2 \sigma_{x,i}^2}},
\eeq
where $R = \log{\lrscale}$, $\mu = \log{\mbhscale}$, $X =
\log{\lxscale}$, and the sum is over each galaxy.  The $\sigma$ terms
are scatter terms that reflect deviation from the plane due to
intrinsic scatter and measurement errors.  This statistic is the same
statistic used by \citet{merlonietal03}.  We considered two cases.
For the first, we assume that the intrinsic scatter is dominant and
isotropic and thus use a total scatter projected in to the $R$
direction: $\sigma_0^2 = \sigma^2_{r,i} + \xi_m^2 \sigma^2_{m,i} +
\xi_x^2 \sigma^2_{x,i}$.  To determine $\sigma_0$, we use a trial
value of $\sigma_0$ and increase the value until the reduced $\chi^2$
is unity after fitting with the new value.  For the second, we use the
measurement errors in \mbh\ and \lx, assumed to be normally
distributed in logarithmic space, for $\sigma_m$ and $\sigma_x$
respectively.  The measurement errors in $\lr$ are likely the
smallest, and thus intrinsic scatter is likely to dominate.  Here we
assume $\sigma_r = \sigma_0$.  In this final case, our fit method is
no longer symmetric, but it includes measurement errors and does not
assume that the intrinsic scatter is isotropic.  Both methods give
nearly identical results, and we report only results from the latter
method, which includes measurement errors.  The errors on fit
parameters come from the formal covariance matrix of the fit.

\subsection{Fundamental Plane Slopes}
\label{fpfit}
\begin{figure*}[t]
\centering
\includegraphics[width=0.45\textwidth]{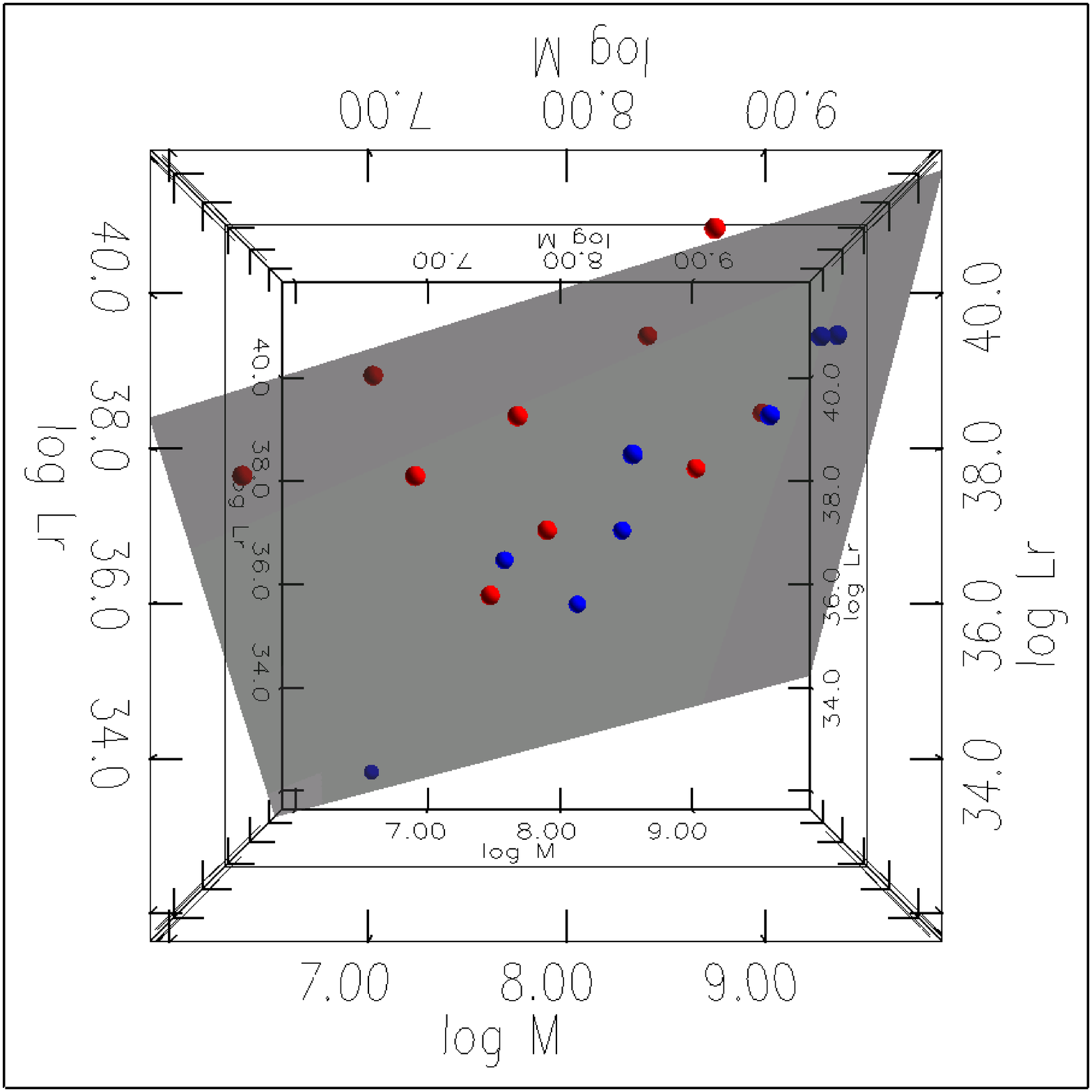}
\includegraphics[width=0.45\textwidth]{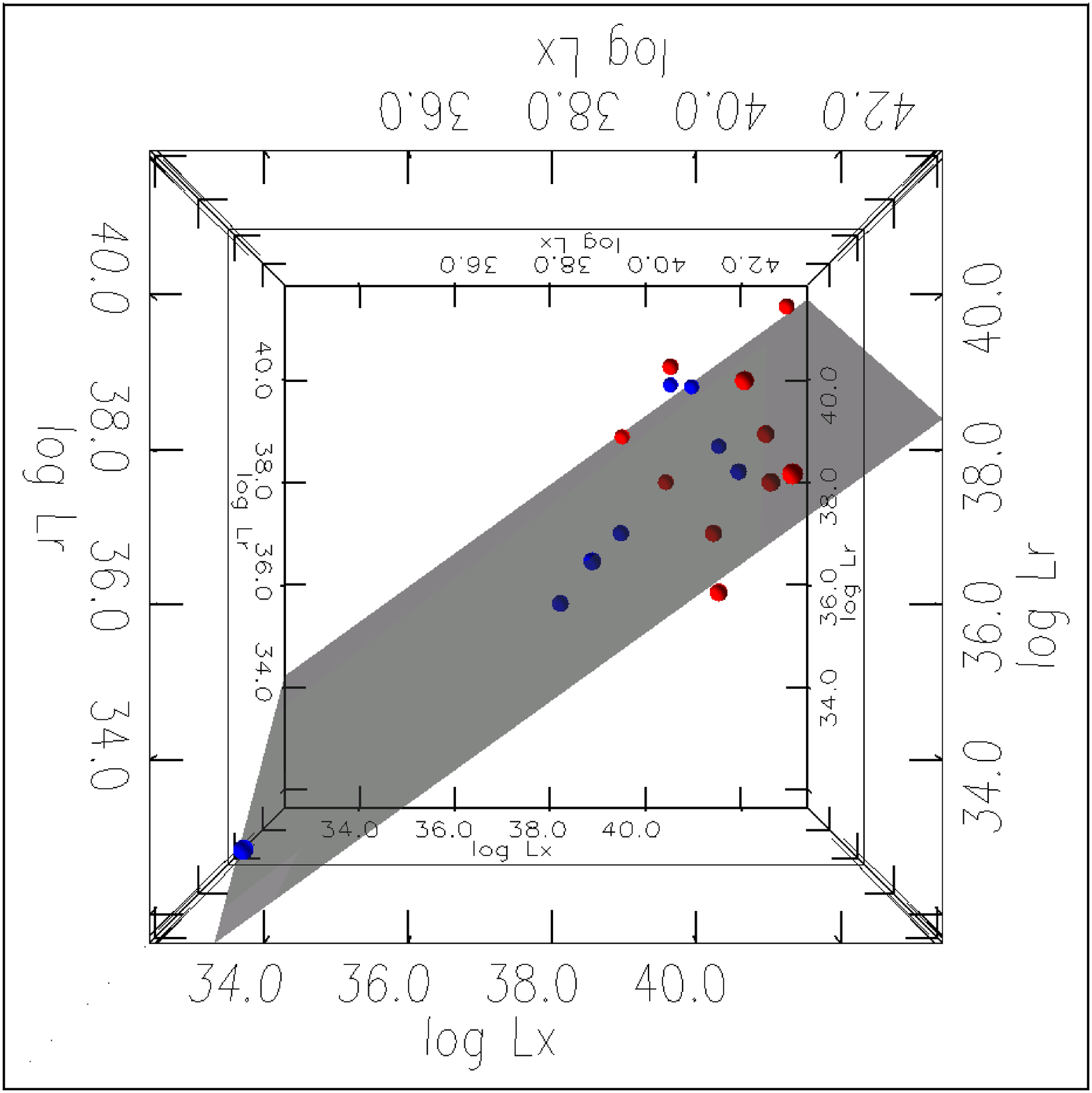}\\
\includegraphics[width=0.45\textwidth]{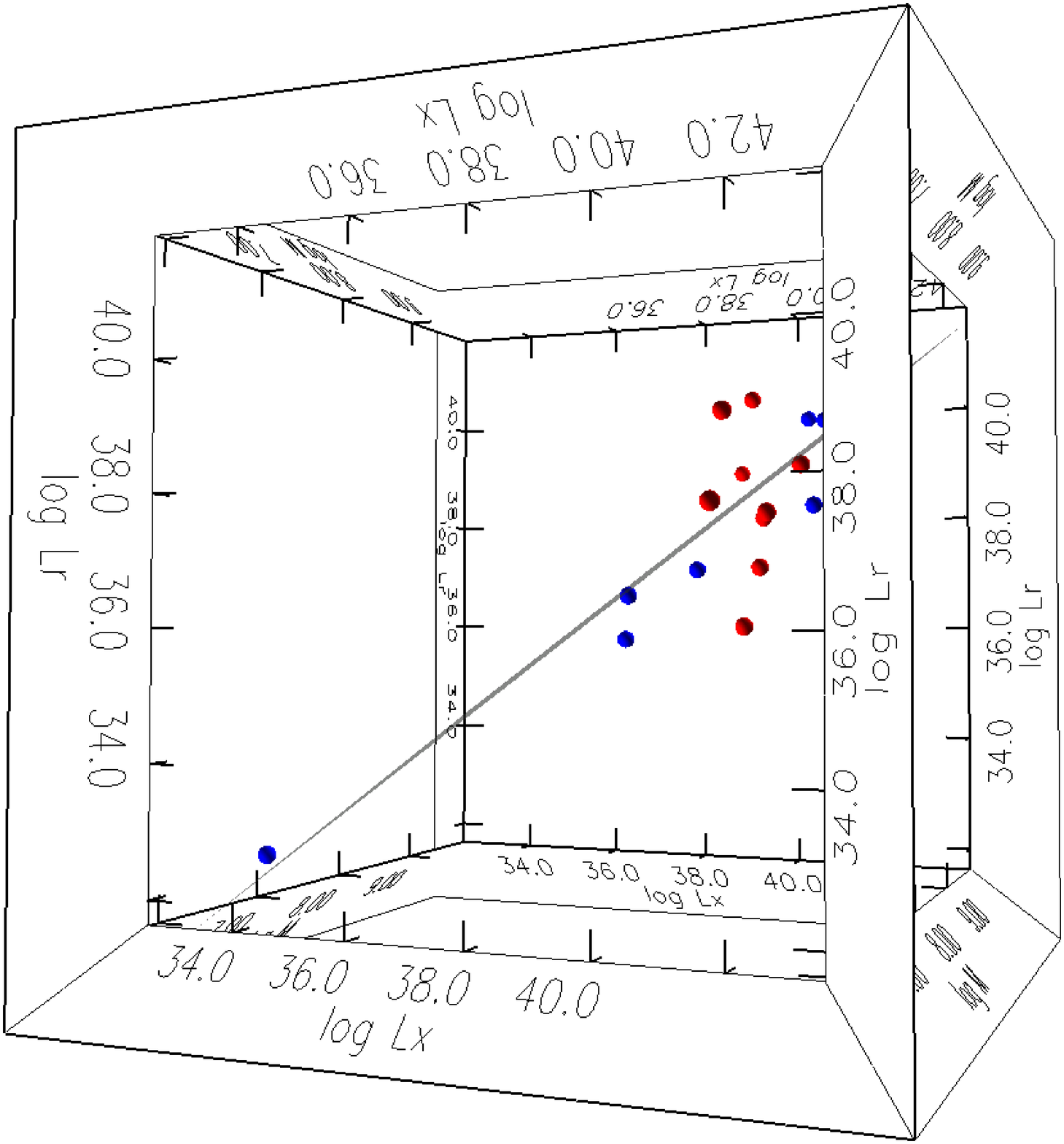}
\includegraphics[width=0.45\textwidth]{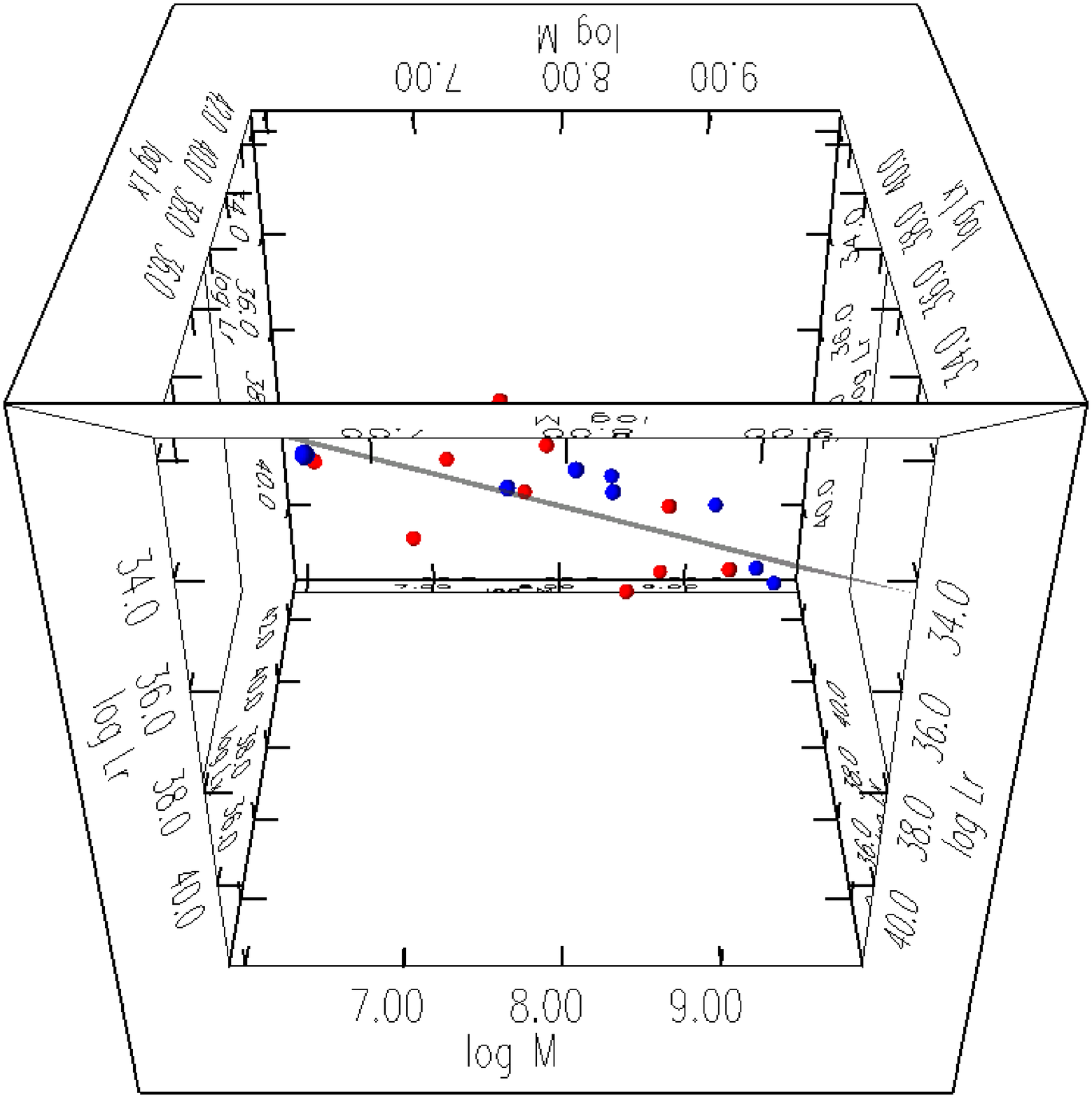}
\caption{Four views of the fundamental plane.  Data are as described
in Sections~\ref{sample} and \ref{xrayanal}.  Red points are galaxies
classified as Seyferts.  Blue points are LLAGNs and LINER galaxies.
The varying views clearly show that as a whole the points lie on a
plane in the dimensions shown.  It is especially clear in the
top-right panel that the LLAGN/LINER subsample appear to lie on a
one-dimensional manifold.}
\label{f:fp3d}
\end{figure*}

\begin{figure}[tbh]
\centering
\includegraphics[width=1.0\columnwidth]{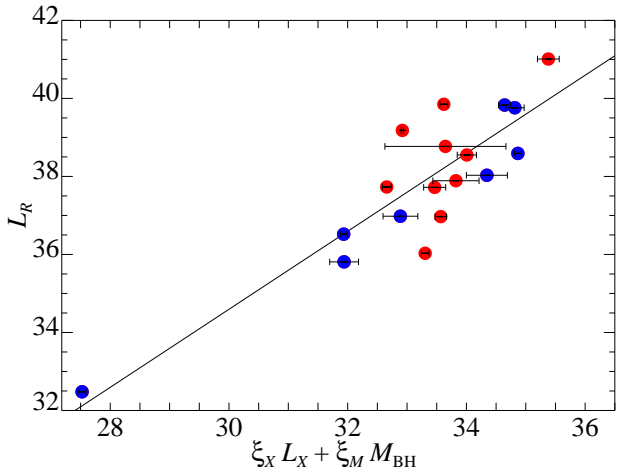}
\caption{The fundamental plane relation.  This figure shows the
edge-on view of our best-fit relation: $\xi_m = 0.78$ and $\xi_x =
0.67$.  Error bars on the $x$-axis are calculated as $\sigma^2_{i} =
\xi^2_m \sigma^2_{m,i} + \xi^2_x \sigma^2_{x,i}$.  This view is
primarily for comparison with \citet{merlonietal03} and with
\citet{fkm04}.  Red circles are Seyferts.  Blue circles are LINERs and
unclassivied LLAGN.}
\label{f:fp}
\end{figure}

Our best-fit relation for the fundamental plane is 
\beqa
\nonumber   R_0 &=& -0.34 \pm  0.24\\
\nonumber \xi_m &=& \phantom{-}0.78 \pm  0.27\\
          \xi_x &=& \phantom{-}0.67 \pm  0.12.
\label{e:bestfp}
\eeqa
The scatter we find in the \lr\ direction is $\sigma_0 = 1.00\
\mathrm{dex}$, equivalent to 0.70 dex normal to the plane.  These
results are consistent with the findings of \citet{merlonietal03} and
of \citet{fkm04}.  We plot several views of the fundamental plane in
Figure~\ref{f:fp3d} and the edge-on view in Figure~\ref{f:fp}.  It is
also interesting to note that for a fixed value of \mbh\ our relation
finds $\lr \propto \lx^{0.67}$, consistent with the findings of
\citet{galloetal03}.

\subsection{\texorpdfstring{\mbh}{Black Hole Mass} as the Dependent Variable}
\label{mpredfit}
\begin{figure*}[t]
\centering
\includegraphics[width=0.45\textwidth]{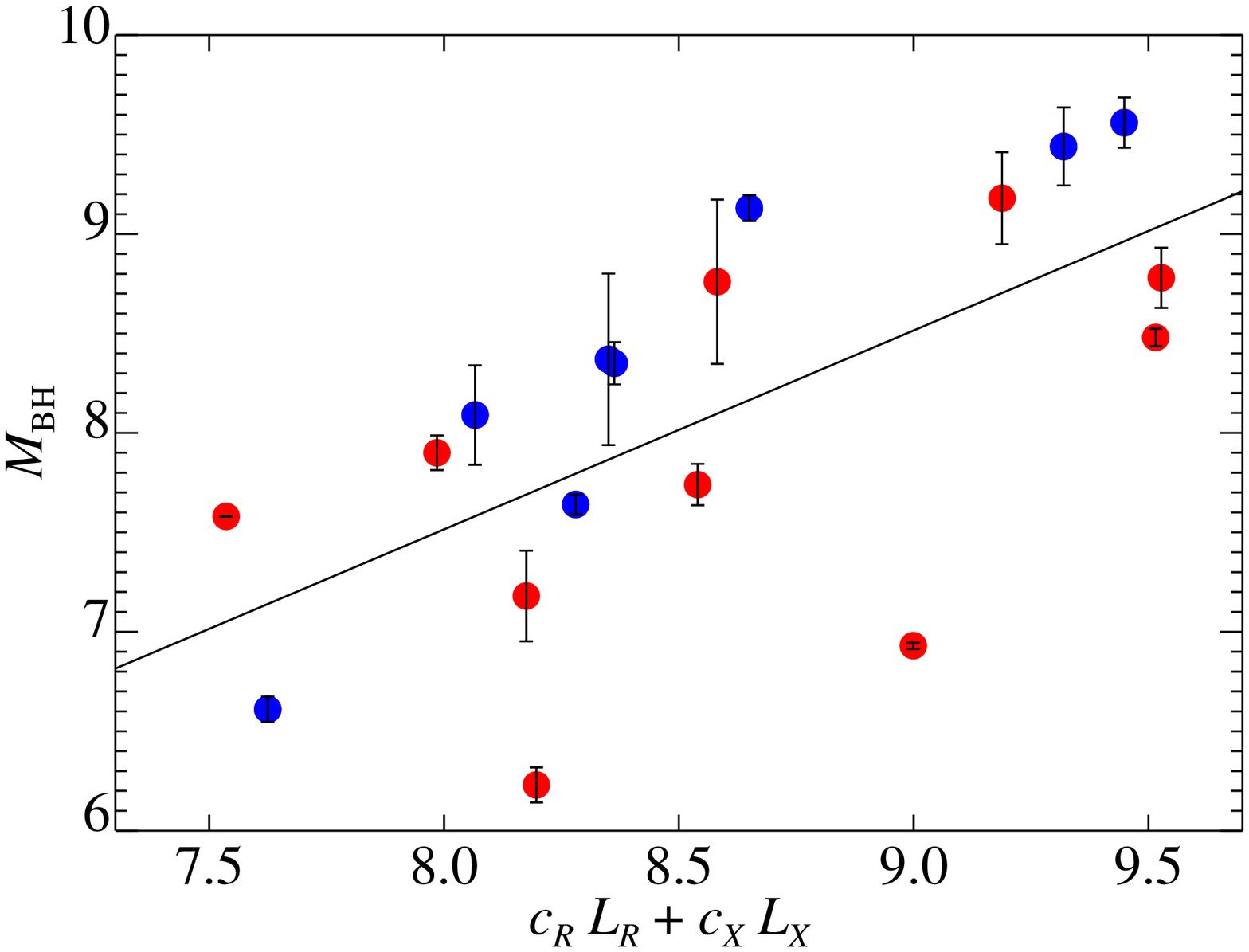}
\includegraphics[width=0.45\textwidth]{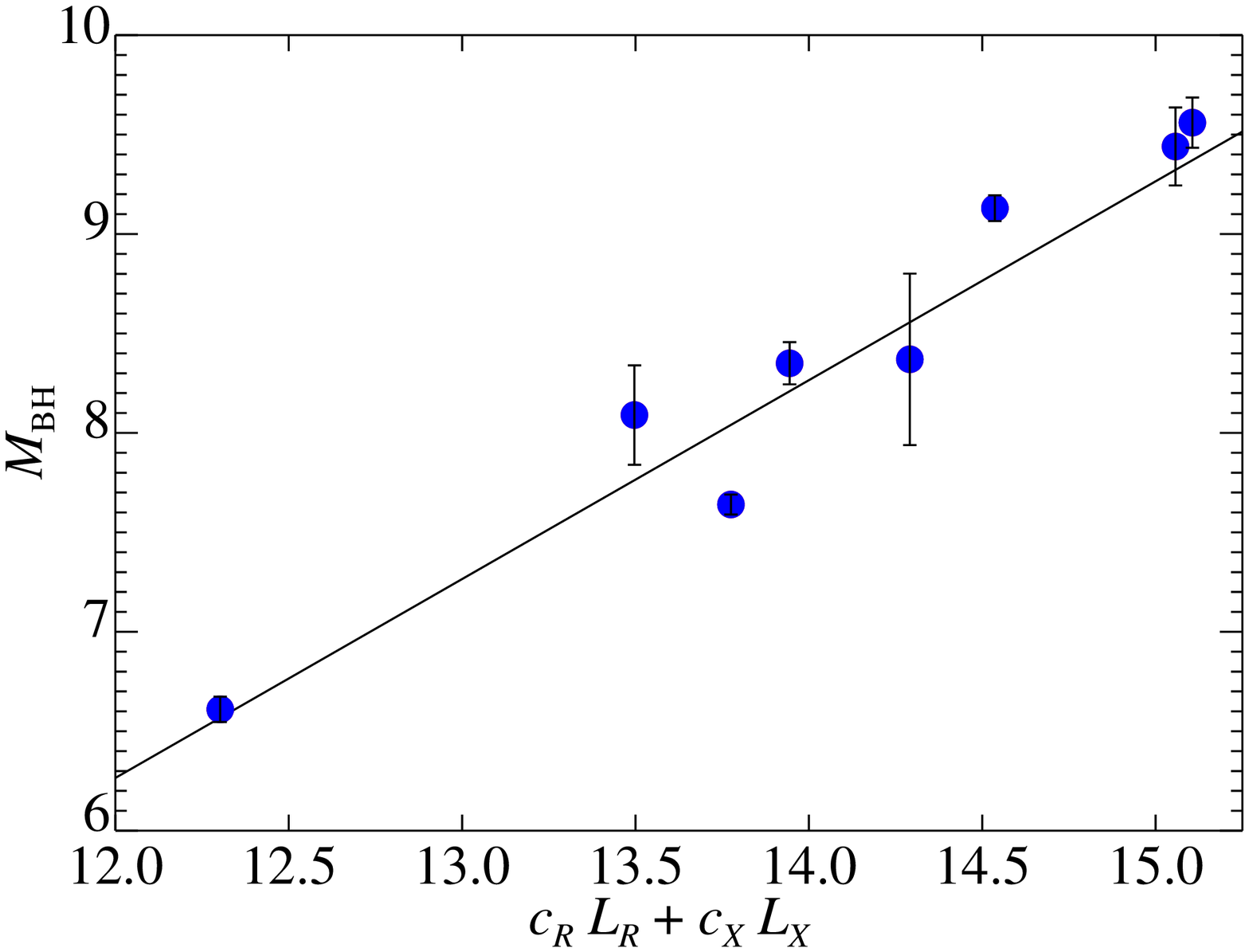}
\caption{Best fit linear regression of \mbh\ on \lr\ and \lx\ for
(left) all galaxies and for (right) LLAGN and LINER galaxies only.
The relation on the right is considerably tighter but may be affected
by the small number of sources.  Red circles are Seyferts.  Blue
circles are LINERs and unclassivied LLAGN.}
\label{f:fpmpredallmextra}
\end{figure*}

We are using black hole masses that have been measured directly.  This
approach allows us to use \lr\ and \lx\ as predictor variables for
\mbh.  We perform a multivariate linear regression on \lr\ and \lx by
assuming a form
\beq
\log\mbhscale = \mu_0 + c_r \log\lrscale + c_x\log\lxscale
\eeq
and minimizing
\beq
\chi^2 = \sum_i\frac{\left(\mu_i - \mu_0 - c_r R - c_x X\right)^2}{\sigma^2_{m,i} + \sigma^2_0},
\eeq
where $\sigma_{m,i}$ is the measurement error in \mbh\ and $\sigma_0$
is an intrinsic scatter term in the $\log(\mbh)$ direction.  As
before, the intrinsic scatter term is increased until the resulting
best fit gives $\chi^2 = 1$.  We find a best-fit relation of
\beqa
\nonumber \mu_0 &=& \phantom{-}0.19 \pm  0.19\\
\nonumber c_r &=& \phantom{-}0.48 \pm  0.16\\
          c_x &=& -0.24 \pm  0.15,
\eeqa
with an intrinsic scatter of $\sigma_0 = 0.77\ \mathrm{dex}$ in the
mass direction.  The intrinsic scatter is larger than other scaling
relations (e.g., $\sigma_0 = 0.44 \pm 0.06$ for the \msigma\ relation
and $\sigma_0 = 0.38 \pm 0.09$ for the \ml\ relation;
\citealt{gultekinetal09b}).  We plot projections of fit in the left
panel of Figure~\ref{f:fpmpredallmextra}.

\section{Discussion}
\label{discussion}

\subsection{Using A Black Hole's Luminosity to Estimate Its Mass}
By using a sample of galaxies that have directly measured black hole
masses, we are able to investigate the correlation between X-ray and
radio luminosity and black hole mass.  The measure of any
correlation's worth as a predictor is the scatter, and we consider the
scatter here.  The scatter in the full relation is considerable
($0.77\ \mathrm{dex} = 5.9$), but it is only a factor of a couple
larger than other scaling relations used to estimate black hole mass.
For example the \msigma\ and \ml\ relations that relate \mbh\ and host
galaxy velocity dispersion and bulge luminosity have intrinsic
scatters of $0.44\ \mathrm{dex} = 2.75$ and $0.38\ \mathrm{dex} =
2.4$, respectively \citep{gultekinetal09b}.

It is worth noting that if we restrict the sample to just black holes
with mass $\mbh > 3 \times 10^7$ or $\mbh > 10^8\ \msun$, the
intrinsic scatter drops to $\sigma_0 = 0.45$ or $0.41$, respectively.  
There are several possible interpretations for the decreased scatter when
restricting the sample by mass.
One possibility is that the requirement of detection in both radio and
X-rays translates to a requirement of high Eddington fraction for
low-mass black holes at a fixed distance.  The mean values of $\fedd$
for the whole sample, for the sample with $\mbh > 3 \times 10^7
\msun$, and for the sample with $\mbh > 10^8 \msun$ are approximately
$6 \times 10^{-5}$, $6 \times 10^{-6}$, and $3 \times 10^{-6}$,
respectively.  It is possible that when sources accrete at a higher
rate, the fundamental plane relation may no longer apply.

Another possible explanation for the smaller scatter in the high-mass
sample is that the low-scatter trend is real, and that the scatter
estimated from the entire sample is skewed by a few data points.  The
most obvious outliers from the left panel of
Figure~\ref{f:fpmpredallmextra} are Circinus and NGC~1068.  If these
two are eliminated, the scatter becomes $\sigma_0 = 0.50\
\mathrm{dex}$.  The derived intrinsic luminosities of these sources
may be difficult to determine because of obscuration.  In these
sources we have a poor view of the central engine and are seeing
reflected, rather than direct X-ray emission
\citep{1996MNRAS.281L..69M,1985ApJ...297..621A}.  If the intrinsic
X-ray luminosity of these sources is higher, then they would lie
closer to the best-fit plane than they do now.

AGN classification for each galaxy of the sample is listed in
Table~\ref{t:data}.  The distinction between Seyferts and LINERs is
judged from the line ratios with the usual diagnostic and division set
so that Seyferts have [OIII]$\lambda$5007/H$\beta > 3.0$
\citep{1987ApJS...63..295V} as a measurement of the level of nuclear
ionization, though there is no obvious transition between the two
classes \citep{2003ApJ...583..159H}.  The physical difference between
LINERs and Seyferts may be that the LINERs lack a ``big blue bump''
and produce a larger partially ionized zone.  The transition in
spectral energy distribution from a Seyfert to a LINER may happen at
low $\fedd$ \citep{2008ARA&A..46..475H}.  The distinction between
Seyfert types is determined by the ratio of broad-line and narrow-line
emission.  LLAGN are defined by having an H$\alpha$ luminosity smaller
than $L(H\alpha) \le 10^{40} \ergs$ \citep{1997ApJS..112..315H}.  The
difference between Seyfert types is understood to be due to differing
viewing angles with respect to an obscuring dusty torus that surrounds
the broad line region (with type~1 unobscured and type~2 completely
obscured).  For a review of the observational differences among the
different classes and the current physical explanations for the
differences see the review by \citet{2008ARA&A..46..475H}.

We may give special consideration to all non-Seyfert AGNs in our
sample.  Since all Seyferts in our sample are at least partially
obscured, obscuration is one potential issue that is addressed.
Obscuration will naturally lead to an underestimate in X-ray
luminosity.  We minimize this by fitting for the absorption across the
0.5--10\ keV band.  Since the softer photons are more readily
absorbed, the shape of the spectrum gives an indication of the level
of absorption.  We also use the hard X-ray flux, which is least
affected by absorption, for our X-ray luminosity.  Nevertheless, the
most heavily obscured sources may be intrinsically brighter than our
fits indicate.  We attempt to isolate this issue below by removing
Compton-thick sources.  In addition to obscuration, as mentioned
above, Seyferts also accrete at higher fractions of Eddington and may
accrete in a mode different from LINERs.  In addition, since Seyferts
are thought to be dominated by thermal output, their radio
luminosities may be poor probes of the power in outflows and thus not
belong on the relation considered here.  Thus, there is a physical
motivation to separate them from the rest of the sample.

When we only use the 8 LINER and unclassified LLAGN sources, our fit
becomes
\beqa
\nonumber \mu_0 &=& \phantom{-}0.70 \pm  0.11\\
\nonumber c_r &=& \phantom{-}0.48 \pm  0.14\\
          c_x &=& -0.07 \pm  0.12,
\eeqa
with a scatter of $\sigma_0 = 0.25$, substantially smaller than other
intrinsic scatter measurements found for this relation and actually
smaller than the scatter in the \msigma\ and \ml\ relations.
\citet{2006A&A...456..439K} similarly found a substantially reduced
scatter in fundamental plane fits to a sample of only stellar-mass
black holes, Sgr~A*, and LLAGNs.  The fit we find is significantly
different from the other fits, notably that it is consistent with no
dependence on X-ray luminosity ($c_x = 0$).  This is at odds with the
findings of \citet{ho02}, who found no dependence of black hole mass
on radio luminosity.  The data do appear to lie on a one-dimensional
manifold in the three-dimensional space considered, but with only 8
data points, the data set is substantially smaller than that of
\citet{ho02}, who also used direct, primary mass measurements in
addition to direct, secondary mass measurements (i.e., reverberation
mapping).

If obscuration, rather than accretion rate or accretion mode, is the
underlying reason for the smaller scatter in the LINER/LLAGN sample,
then we should see similar results when omitting sources that are
Compton thick ($n_H \ge \sigma_T^{-1} = 1.5 \times 10^{24}\
\mathrm{cm^{-2}}$).  Compton-thick sources will be heavily obscured
and the intrinsic luminosities may be much higher than the observed
flux would imply \citep{2002ApJ...573L..81L, 2006ApJ...648..111L}.  If
we conservatively omit the sources from Table~\ref{t:xray} intrinsic
absorption larger than $10^{24}\ \mathrm{cm^{-2}}$ (NGC~3031,
NGC~4374, and NGC~6251) as well as the sources determined to be
Compton thick from Fe K$\alpha$ modeling \citep[Circinus and
NGC~1068;][]{2002ApJ...573L..81L, 2006ApJ...648..111L}, we obtain
\beqa
\nonumber \mu_0 &=& \phantom{-}0.40 \pm  0.16\\
\nonumber c_r &=& \phantom{-}0.46 \pm  0.13\\
          c_x &=& -0.14 \pm  0.12,
\eeqa
with a scatter of $\sigma_0 = 0.53$.  This result is consistent with
the Seyfertless sample at about the 1$\sigma$ level, though with a
larger scatter.

\subsection{Sgr~A*}
\label{sgrastar}
Sgr~A*, the central black hole in the Galaxy, is a unique source in
many ways.  Its extremely low accretion rate ($\lx / \ledd \approx
4\times10^{-12}$) is two orders of magnitude below the next lowest in
our sample.  An analog to Sgr~A* could not be observed outside of the
local group.

When using only the two nearby super-massive black holes with
extremely well-determined mass and distance (Sgr~A* and NGC~4258) and
the X-ray binary in which the correlation extends over several orders
of magnitude (GX 339$-$4), the best-fit fundamental plane relation
changes so that Sgr~A* is under-luminous in X-rays during quiescence
by at least 2 orders of magnitude \citep{2005ApJ...618L.103M}.  Such a
break from the correlation \citep[also seen in some X-ray binaries as
they rise out of quiescence][]{2009arXiv0909.3283C} may indicate that,
during quiescence at least, Sgr~A* is accreting in a different mode
than the correlation sources.  If such an extremely low accretion rate
is in a different category from the rest of the objects, then it makes
sense to exclude Sgr~A* from the sample, in which case our best fit
becomes
\beqa
\nonumber R_0 &=& -0.46 \pm  0.30\\
\nonumber \xi_m &=& \phantom{-}0.84 \pm  0.29\\
          \xi_x &=& \phantom{-}0.80 \pm  0.23,
\eeqa
with an intrinsic scatter of $\sigma_0 = 1.02$, which is not a
significantly different fit.  

\subsection{Stellar-mass Sources}
\label{stellarmass}
Our initial sample includes only the supermassive black holes in
galactic centers.  There are, however, several Galactic stellar-mass
black holes with dynamically measured masses.  If accretion onto black
holes is driven by the same physical processes at all mass scales,
then the stellar-mass sources should obey the same relation, which is
what \citet{merlonietal03} and \citet{fkm04} found.  So while our
focus has been on super-massive black holes, we may revisit our
calculations with the sample of stellar-mass black holes given in
Table~\ref{t:stellardata}.  This sample was selected from stellar-mass
black holes with dynamically determined masses with simultaneous X-ray
and radio data.  In addition to the sources listed, there were two
stellar mass black holes that had adequate data (4U~1543$-$475 and
GRO~J1655$-$40) but whose jets may not be in a steady state and thus
skewing the relation.

The stellar-mass systems, with the possible exception of GRS~1915, are
in the low/hard state, which is characterized by a hard X-ray photon
index ($1.4 < \Gamma < 2.1$), a small ratio of unabsorbed disk flux to
total unabsorbed flux \citep[$f < 0.2$;][]{2006ARA&A..44...49R} and is
usually seen at low Eddington rates.  This state is also typically
associated with a steady radio jet whereas jets in the high/soft state
are quenched \citep{2001MNRAS.322...31F}.  By requiring radio
emission, we essentially require a low/hard state.  If such a state
can be extended to SMBH sources, it would naturally compare with the
similarly low Eddington rates in LLAGNs in which jet emission is more
prominent compared to Seyferts.  The mapping of X-ray binary states to
accreting SMBHs is complicated by the fact that no comparable
transitions are seen in SMBHs.

These three accreting black holes have masses measured from period
measurements of the donor star's orbit.  The mass of the donor star is
estimated based on spectral type, and the inclination of the orbit for
systems such as these is generally derived from modeling the star's
change in flux, assumed to be from the change in viewing angle of a
tear-drop-shaped object (ellipsoidal modulation).  For two of the
three stellar-mass sources we are using, however, the inclination is
constrained by other means.  For GRS 1915+105 the inclination is
constrained from the apparent superluminal motion of ejected jet
material that is assumed to be perpendicular to the orbital plane
based on the lack of observed precession
\citep{1994Natur.371...46M,2001A&A...373L..37G}.  For Cyg~X-1, the
inclination has been estimated in several ways, including UV line
modeling and X-ray polarization \citep[][and references
therein]{1987ApJ...321..425N}.

The luminosity data from each source is simultaneous, which is
important for these highly variable sources.  For two of the sources,
we use two sets of simultaneous observations.  Using more than one
observation of a particular source in the fit over-weights that source
and will skew the fit if it is atypical.  Under the assumption that
each source belongs in the fit in all of the epochs used, however,
they provide valuable extra information of possible accretion states
in the same relation.

\begin{deluxetable}{lrr@{$\pm$}lrr@{$\pm$}lr}
\tablecaption{Stellar-Mass Black Hole Data}
\tablewidth{0pt}
\tablehead{
\colhead{Name} & 
\colhead{$D$} & 
\multicolumn{2}{c}{$\log\left(\mbh\right)$} & 
\colhead{$\log\left(L_R\right)$} & 
\multicolumn{2}{c}{$\log\left(L_X\right)$} &
\colhead{Refs.}}
\startdata
GRS 1915+105 & 11    & 1.15 & 0.13 & 30.64\tablenotemark{a} & 38.06 & 0.06 & 1,2,3,3\\
\dots        & 11    & 1.15 & 0.13 & 30.90\tablenotemark{a} & 38.69 & 0.06 & 1,2,3,3\\
V404 Cyg     &  3    & 1.08 & 0.07 & 28.30 & 33.07 & 0.24 & 4,4,5,6\\
Cygnus X-1   &  2.5  & 1.00 & 0.24\tablenotemark{b} & 29.91\tablenotemark{c} & 36.71 & 0.18\tablenotemark{d} & 7,8,9\\
\dots        &  2.5  & 1.00 & 0.24\tablenotemark{b} & 29.84\tablenotemark{c} & 36.77 & 0.18\tablenotemark{d} & 7,8,9
\enddata
\label{t:stellardata}
\tablecomments{Stellar-mass black hole data used in
section~\ref{stellarmass}.  Distances are given in units of
$\mathrm{kpc}$.  Black hole masses are in solar units.  Radio and
X-ray luminosities are in units of $\mathrm{erg\ s^{-1}}$.  All values
are scaled to the distances given.  The sources were in low/hard state
for the epochs listed with the exception of GRS~1915, which may be in
a plateau state \citep{2001ApJ...556..515M}.  The numbers in the
reference column give the number of the original reference for the
distance, mass, radio luminosity, and X-ray luminosity, respectively.
X-ray luminosities have been converted to the $E = 2$--$10\
\mathrm{keV}$ band.}
\tablenotetext{a}{Interpolated from $\nu = 2.25\ \mathrm{GHz}$ and
$\nu = 8.3\ \mathrm{GHz}$ data.}
\tablenotetext{b}{Mass uncertainty was estimated from the range of
values found in the literature \citep{2006csxs.book..157M}.}
\tablenotetext{c}{Extrapolated from $\nu = 8.4\ \mathrm{GHz}$ assuming
constant $\nu{F_\nu}$.}
\tablenotetext{d}{Data come from Rossi X-ray Timing Explorer (RXTE)
All-Sky Monitor (ASM) assuming a standard spectral form.}
\tablerefs{
(1) \citealt{1999MNRAS.304..865F};
(2) \citealt{2001Natur.414..522G};
(3) \citealt{2001ApJ...556..515M};
(4) \citealt{1994MNRAS.271L..10S};
(5) \citealt{2005MNRAS.356.1017G};
(6) \citealt{2007ApJ...667..427B};
(7) \citealt{1973ApJ...185L.117B};
(8) \citealt{1995A&A...297..556H};
(9) \citealt{2001MNRAS.327.1273S}.
}
\end{deluxetable}

The results of our fundamental plane fits become:
\beqa
\nonumber   R_0 &=& -0.33 \pm 0.21\\
\nonumber \xi_m &=& \phantom{-}0.82 \pm 0.08\\
          \xi_x &=& \phantom{-}0.62 \pm 0.10,
\eeqa
with an intrinsic scatter of $\sigma_0 = 0.88$.  The uncertainties in
slopes have decreased because of the increased range in the values
present, especially for $\xi_m$.  It is interesting to note that while
the best-fit parameters do not significantly change from our fits to
central black holes, the intrinsic scatter does.  This decrease can be
attributed to the fact that these sources lie closer to the plane.  It
is also worth noting that the fits do not change even though two of
the stellar-mass sources are accreting at a much higher fraction of
Eddington than the supermassive sources.  GRS 1915+105 is accreting at
$\fedd \approx 0.06$ to $0.3$, and Cygnus X-1 at $\fedd \approx 0.004$
to $0.005$, whereas all of the supermassive sources are accreting at
$\fedd < 0.001$ (Fig.~\ref{f:fedd}).

It should be noted that there are different systematic errors in the
stellar-mass and central black holes.  The mass measurements are from
completely different methods.  The X-ray extragalactic sources may be
contaminated from point sources and may be more heavily obscured than
the stellar-mass sources.  The extragalactic sources may also be
contaminated by supernova remnants along the line of sight, though
this can be mitigated by going to higher frequencies.  Stellar mass
uncertainties are dominated by uncertainties in distance, inclination,
and light from accretion \citep[see][]{2008MNRAS.387..788R}.

\subsection{Future Work}
\label{future}
In this paper, we have only included the 18 black holes with measured
masses, radio fluxes, and X-ray fluxes.  This sample makes up slightly
more than one third of the entire sample of black holes with measured
masses.  There are 11 without nuclear radio data or only with upper
limits on one or more of these quantities.  There are a further 16
sources with no \emph{Chandra} X-ray fluxes measured because either
there are no \emph{Chandra} data or merely insufficient data.  Many of
the sources have masses $M < 10^8\,\msun$.  By completing the sample
of \msigma\ black holes with further X-ray and radio observations, the
increased number of data points should be especially helpful in
determining whether the large scatter at the low-mass end and the
small scatter at the high-mass end are actual differences or just
artifacts of a few outliers.

Another place for future work is in understanding the apparent special
place that Seyfert galaxies occupy in the fundamental plane.  If one
were to na\"{\i}vely assign accretion states used for stellar-mass
black holes to Seyfert galaxies, they would be considered in the
thermally dominant/high--soft state.  For stellar-mass black holes in
this state, jets are not measured.  That the Seyfert galaxies are an
apparent source of scatter in the relation may be an indication that
they are diverging away from the fundamental plane relation.  To
better understand the differences between Seyfert galaxies and the
other sources, a future theoretical work will consider just these
types of sources, including physical modeling of the data sets
presented here.

\section{Conclusions}
\label{concl}
In this paper we analyze the relationship among X-ray luminosity,
radio luminosity, and the mass of a black hole.  Distinct from
previous studies of this relationship, we use only black hole masses
that have been dynamically measured.  Because of the relatively small
distances to the objects in this sample, we avoid potential biases
arising from flux limited samples.  Using the most recent compilation
of black hole masses, we analyzed archival \emph{Chandra} data to get
nuclear X-ray luminosities in the $E = 2$--$10\ \mathrm{keV}$ band.
We combined this with $\nu = 5\ \mathrm{GHz}$ radio luminosities found
in the literature and fit a relation of the form $\lrscale = R_0 + \xi_m
\log\mbhscale + \xi_x \log\lxscale$ to find
\beqa
\nonumber   R_0 &=& -0.34 \pm  0.24\\
\nonumber \xi_m &=& \phantom{-}0.78 \pm  0.27\\
          \xi_x &=& \phantom{-}0.67 \pm  0.12
\label{e:bestfp2}
\eeqa
with a scatter of $\sigma = 1.00$ in the $\log\lr$ direction,
consistent with previous work.  We also fit a relation to be used as
an estimation for black hole mass based on observations of \lx\ and
\lr of the form:
\beq
\log\mbhscale = \mu_0 + c_r \log\lrscale + c_x\log\lxscale,
\eeq
finding
\beqa
\nonumber \mu_0 &=& \phantom{-}0.19 \pm  0.19\\
\nonumber c_r &=& \phantom{-}0.48 \pm  0.16\\
          c_x &=& -0.24 \pm  0.15,
\eeqa
with an intrinsic scatter of $\sigma_0 = 0.77$ in the $\log\mbh$
direction.  This intrinsic scatter is larger than other scaling
relations involving \mbh, but decreases considerably when only using
the most massive black holes or when eliminating obscured central
engines from the sample.  Both of these issues require further
investigation and could be answered by completing the sample with more
\emph{Chandra} observations.

\acknowledgements We thank the anonymous referee for useful comments
that have improved this paper.  KG thanks Fill Humphrey and Tom
Maccarone for helpful comments.
EMC gratefully acknowledges support provided by the National
Aeronautics and Space Administration (NASA) through the Chandra
Fellowship Program, grant number PF8-90052.
This work made use of the VizieR catalog access tool, CDS,
Strasbourg, France; data products from the Two Micron All Sky Survey
(2MASS), which is a joint project of the University of Massachusetts
and the Infrared Processing and Analysis Center/California Institute
of Technology, funded by NASA and the National Science Foundation;
NASA's Astrophysics Data System (ADS); and the NASA/IPAC Extragalactic
Database (NED), which is operated by the Jet Propulsion Laboratory,
California Institute of Technology, under contract with NASA.
Three-dimensional visualization was made possible by the S2PLOT
programming library described in \citet{2006PASA...23...82B}.

\bibliographystyle{apjads}
\bibliography{gultekin}

\clearpage
    \ifthenelse{\value{emulateapj} = 1}
{
\begin{landscape}
    \tabletypesize{\scriptsize}
    \def\arraystretch{1.200}
}
{

}
    \begin{deluxetable*}{lrrrrrrrrrrrrr}
    \ifthenelse{\value{emulateapj} = 1}
{}
{
    \tabletypesize{\scriptsize}
    \rotate
    \setlength{\tabcolsep}{0.05in}
}
    \tablecaption{Summary of \emph{Chandra} Spectral Fits}
    \tablewidth{0pt}
    \tablehead{
    \colhead{Galaxy} & \colhead{Obs. ID} & \colhead{Exp.} & \colhead{$\chi^2 / \nu$} & \colhead{Galactic absorption} & \multicolumn{2}{c}{Intrinsic absorption} & \multicolumn{2}{c}{Power-law} & \multicolumn{2}{c}{APEC} & \multicolumn{3}{c}{Gaussian}\\
           &       & [ks] & & \colhead{$n_H$ [$\mathrm{cm}^{-2}$]}               & \colhead{$n_H$ [$\mathrm{cm}^{-2}$]} & \colhead{$f_\mathrm{cov}$} & \colhead{$\Gamma$} & \colhead{$A_\mathrm{pl}$} & \colhead{$kT_\mathrm{APEC}$ [$\mathrm{keV}$]} & \colhead{$A_\mathrm{APEC}$} & \colhead{$E_\mathrm{line}$ [$\mathrm{keV}$]} & \colhead{$\sigma_\mathrm{line}$ [$\mathrm{keV}$]} & \colhead{$A_\mathrm{line}$}
    }
    \startdata
Circinus & 356 & 24.7 & $261.9 / 166$ &
$1.88^{+0.05}_{-0.07}\times10^{22}$ & \dots & \dots & 
$-1.39_{-0.14}^{+0.10}$ & 
$1.40^{+0.12}_{-0.26}\times10^{-4}$ & 
$1.00_{-0.04}^{+0.05}$ &
$2.96^{+0.18}_{-0.25}\times10^{-2}$ & 
$6.40_{-0.00}^{+0.00}$ & 
$3.41^{+0.82}_{-0.96}\times10^{-2}$ & 
$2.87^{+0.13}_{-0.15}\times10^{-3}$\\
CygnusA & 1707 & 9.2 & $143.6 / 112$ &
$1.99^{+0.57}_{-0.47}\times10^{21}$ & 
$1.47^{+0.07}_{-0.07}\times10^{23}$ & 
$0.98_{-0.00}^{+0.00}$ & 
$1.34_{-0.08}^{+0.09}$ & 
$3.04^{+0.51}_{-0.41}\times10^{-3}$ & \dots & \dots & 
$6.07_{-0.02}^{+0.02}$ & 
$7.06^{+2.22}_{-2.61}\times10^{-2}$ & 
$8.47^{+1.41}_{-1.35}\times10^{-5}$\\
IC1459 & 2196 & 58.8 & $189.5 / 178$ &
$2.13^{+0.12}_{-0.12}\times10^{21}$ & \dots & \dots &
$1.96_{-0.04}^{+0.04}$ & 
$2.35^{+0.09}_{-0.08}\times10^{-4}$ & \dots & \dots & \dots & \dots & \dots\\
IC4296 & 3394 & 24.8 & $85.6 / 74$ &
$1.40^{+0.27}_{-0.22}\times10^{21}$ & \dots & \dots & 
$0.80_{-0.08}^{+0.08}$ & 
$3.24^{+0.36}_{-0.33}\times10^{-5}$ & 
$0.55_{-0.02}^{+0.02}$ & 
$1.20^{+0.14}_{-0.14}\times10^{-4}$ & \dots & \dots & \dots\\
N0221 & 5690 & 113.0 & $19.5 / 22$ &
$8.33^{+30.39}_{-8.33}\times10^{19}$ & \dots & \dots &
$2.01_{-0.11}^{+0.16}$ & 
$6.71^{+1.00}_{-0.49}\times10^{-6}$ & \dots & \dots & \dots & \dots & \dots\\
N0821 & 6313 & 49.5 & \dots &
$8.53^{+14.85}_{-8.53}\times10^{20}$ & \dots & \dots &
$2.00_{-0.48}^{+0.56}$ & 
$1.38^{+0.91}_{-0.47}\times10^{-6}$ & \dots & \dots & \dots & \dots & \dots\\
N1023 & 8464 & 47.6 & $6.4 / 17$ &
$1.46^{+0.43}_{-0.45}\times10^{21}$ & \dots & \dots &
$2.15_{-0.15}^{+0.14}$ & 
$1.80^{+0.34}_{-0.28}\times10^{-5}$ & \dots & \dots & \dots & \dots & \dots\\
N1068 & 344 & 47.4 & $217.8 / 125$ &
$1.32^{+0.10}_{-0.15}\times10^{21}$ & \dots & \dots & 
$3.48_{-0.09}^{+0.09}$ & 
$3.26^{+0.23}_{-0.20}\times10^{-4}$ & 
$0.80_{-0.02}^{+0.02}$ & 
$7.87^{+0.46}_{-0.70}\times10^{-5}$ & \dots & \dots & \dots\\
N1399\tablenotemark{a} & 319 & 57.4 & $17.7 / 12$ &
$4.71^{+5.79}_{-2.71}\times10^{21}$ & \dots & \dots &
$4.62_{-1.98}^{+3.64}$ & 
$<5.68\times10^{-5}$ & \dots & \dots & \dots & \dots & \dots\\
N2787 & 4689 & 30.9 & $ 18.6 / 21 $ &
$1.27^{+0.40}_{-0.41}\times10^{21}$ & \dots & \dots &
$2.20_{-0.16}^{+0.15}$ & 
$3.53^{+0.57}_{-0.48}\times10^{-5}$ & \dots & \dots & \dots & \dots & \dots\\
N3031 & 6897 & 14.8 & $119.9 / 90$ &
$<1.73\times10^{20}$ & 
$1.14^{+0.15}_{-0.31}\times10^{24}$ & 
$0.85_{-0.10}^{+0.07}$ & 
$1.78_{-0.04}^{+0.06}$ & 
$9.62^{+8.42}_{-3.78}\times10^{-3}$ & \dots & \dots & \dots & \dots & \dots\\
N3115 & 2040 & 37.0 & $5.7 / 3$ &
$1.44^{+0.88}_{-1.13}\times10^{21}$ & \dots & \dots &
$2.35_{-0.68}^{+0.89}$ & 
$5.75^{+4.17}_{-2.19}\times10^{-6}$ & \dots & \dots & \dots & \dots & \dots\\
N3227 & 860 & 49.3 & $316.5 / 233$ &
$<1.29\times10^{20}$ & 
$8.43^{+1.92}_{-1.93}\times10^{21}$ & 
$0.53_{-0.08}^{+0.04}$ & 
$0.69_{-0.08}^{+0.05}$ & 
$4.83^{+0.38}_{-0.54}\times10^{-4}$ & \dots & \dots & 
$6.24_{-0.02}^{+0.02}$ & 
$1.81^{+375.39}_{-1.81}\times10^{-4}$ & 
$1.35^{+0.31}_{-0.31}\times10^{-5}$\\
N3245 & 2926 & 9.6 & \dots &
$1.62^{+1.29}_{-1.18}\times10^{21}$ & \dots & \dots &
$1.90_{-0.41}^{+0.44}$ & 
$1.13^{+0.59}_{-0.36}\times10^{-5}$ & \dots & \dots & \dots & \dots & \dots\\
N3377 & 2934 & 39.6 & $1.7 / 3$ &
$2.94^{+0.92}_{-1.17}\times10^{21}$ & \dots & \dots &
$3.14_{-0.65}^{+0.75}$ & 
$1.16^{+0.71}_{-0.42}\times10^{-5}$ & \dots & \dots & \dots & \dots & \dots\\
N3379 & 7076 & 69.3 & $3.2 / 4$ &
$8.19^{+6.83}_{-8.19}\times10^{20}$ & \dots & \dots &
$2.05_{-0.46}^{+0.39}$ & 
$3.82^{+1.78}_{-1.09}\times10^{-6}$ & \dots & \dots & \dots & \dots & \dots\\
N3384\tablenotemark{a} & 4692 & 9.9 & \dots &
$3.15^{+1.75}_{-1.87}\times10^{21}$ & \dots & \dots &
$3.25_{-0.83}^{+0.78}$ & 
$1.80^{+1.32}_{-0.83}\times10^{-5}$ & \dots & \dots & \dots & \dots & \dots\\
N3585 & 2078 & 35.3 & $19.2 / 6$ &
$9.69^{+0.00}_{-0.00}\times10^{20}$ & \dots & \dots &
$2.09_{-0.00}^{+0.00}$ & 
$8.02^{+0.00}_{-0.00}\times10^{-6}$ & \dots & \dots & \dots & \dots & \dots\\
N3607\tablenotemark{a} & 2073 & 38.5 & \dots &
$7.90^{+0.34}_{-0.30}\times10^{21}$ & \dots & \dots &
$7.70_{-1.81}^{+2.23}$ & 
$<6.29\times10^{-5}$ & \dots & \dots & \dots & \dots & \dots\\
N3608\tablenotemark{a} & 2073 & 38.5 & $0.28 / 2$ &
$5.03^{+3.39}_{-1.57}\times10^{21}$ & \dots & \dots &
$5.59_{-1.61}^{+2.28}$ & 
$<1.02\times10^{-4}$ & \dots & \dots & \dots & \dots & \dots\\
N3998 & 6781 & 13.6 & $421.7 / 297$ &
$5.28^{+6.52}_{-5.28}\times10^{19}$ & \dots & \dots &
$1.37_{-0.02}^{+0.02}$ & 
$1.52^{+0.03}_{-0.03}\times10^{-3}$ & \dots & \dots & \dots & \dots & \dots\\
N4026\tablenotemark{a} & 6782 & 13.8 & \dots &
$3.46^{+3.01}_{-2.37}\times10^{21}$ & \dots & \dots &
$3.47_{-1.13}^{+1.54}$ & 
$7.06^{+10.74}_{-7.06}\times10^{-6}$ & \dots & \dots & \dots & \dots & \dots\\
N4151 & 335 & 47.4 & $366.7 / 253$ &
$<5.12\times10^{21}$ & 
$7.51^{+26.17}_{-7.51}\times10^{21}$ & 
$0.05_{-0.05}^{+0.95}$ & 
$-0.92_{-0.06}^{+0.03}$ & 
$4.71^{+0.33}_{-0.41}\times10^{-5}$ & 
$0.61_{-0.03}^{+0.03}$ & 
$1.06^{+0.00}_{-0.00}\times10^{-4}$ & 
$6.40_{-0.01}^{+0.01}$ & 
$7.36^{+3886.77}_{-7.36}\times10^{-5}$ & 
$3.49^{+0.55}_{-0.51}\times10^{-5}$\\
N4258 & 2340 & 6.9 & $67.3 / 69$ &
$2.69^{+8.33}_{-1.77}\times10^{20}$ & 
$6.68^{+0.56}_{-0.48}\times10^{22}$ & 
$0.99_{-0.00}^{+0.00}$ & 
$1.45_{-0.14}^{+0.17}$ & 
$1.84^{+0.58}_{-0.38}\times10^{-3}$ & \dots & \dots & \dots & \dots & \dots\\
N4261\tablenotemark{a} & 9569 & 101.0 & $185.8 / 169$ &
$9.48^{+1.38}_{-1.35}\times10^{20}$ & 
$5.32^{+0.79}_{-0.81}\times10^{20}$ & $0.90^{+0.02}_{-0.03}$ &
$1.35_{-0.10}^{+0.04}$ & 
$1.14^{+0.21}_{-0.24}\times10^{-4}$ & 
$0.58_{-0.01}^{+0.01}$ & 
$1.38^{+0.08}_{-0.08}\times10^{-4}$ & \dots & \dots & \dots\\
N4303 & 2149 & 28.0 & $19.2 / 8$ &
$3.09^{+5.56}_{-3.09}\times10^{20}$ & \dots & \dots &
$2.14_{-0.28}^{+0.38}$ & 
$9.92^{+3.19}_{-1.67}\times10^{-6}$ & \dots & \dots & \dots & \dots & \dots\\
N4342 & 4687 & 38.3 & $6.2 / 5$ &
$<5.39\times10^{20}$ & \dots & \dots &
$1.44_{-0.17}^{+0.29}$ & 
$5.73^{+1.88}_{-0.52}\times10^{-6}$ & \dots & \dots & \dots & \dots & \dots\\
N4374 & 803 & 28.5 & $18.2 / 28$ &
$2.01^{+0.45}_{-0.42}\times10^{21}$ & 
$3.74^{+399.93}_{-0.39}\times10^{24}$ & 
$1.00_{-0.73}^{+0.00}$ & 
$2.20_{-0.17}^{+0.18}$ & 
$4.29^{+0.21}_{-0.33}\times10^{-2}$ & \dots & \dots & \dots & \dots & \dots\\
N4459\tablenotemark{a} & 2927 & 9.8 & \dots &
$2.55^{+1.34}_{-1.37}\times10^{21}$ & \dots & \dots &
$3.22_{-0.64}^{+0.65}$ & 
$2.06^{+1.12}_{-0.78}\times10^{-5}$ & \dots & \dots & \dots & \dots & \dots\\
N4473\tablenotemark{a} & 4688 & 29.6 & \dots &
$9.84^{+14.56}_{-9.84}\times10^{20}$ & \dots & \dots &
$2.33_{-0.54}^{+0.65}$ & 
$3.85^{+2.47}_{-1.34}\times10^{-6}$ & \dots & \dots & \dots & \dots & \dots\\
N4486 & 2707 & 98.7 & $344.6 / 216$ &
$5.91^{+13.02}_{-5.91}\times10^{19}$ & \dots & \dots &
$0.81_{-0.03}^{+0.03}$ & 
$4.71^{+0.19}_{-0.16}\times10^{-5}$ & \dots & \dots & \dots & \dots & \dots\\
N4486A\tablenotemark{a} & 8063 & 5.1 & \dots &
$7.95^{+7.48}_{-4.95}\times10^{21}$ & \dots & \dots &
$6.14_{-2.60}^{+6.14}$ & 
$<4.81\times10^{-4}$ & \dots & \dots & \dots & \dots & \dots\\
N4564\tablenotemark{a} & 4008 & 17.9 & \dots &
$1.25^{+10.64}_{-1.25}\times10^{20}$ & \dots & \dots &
$1.93_{-0.25}^{+0.47}$ & 
$6.53^{+2.99}_{-1.05}\times10^{-6}$ & \dots & \dots & \dots & \dots & \dots\\
N4594 & 1586 & 18.5 & $110.4 / 102$ &
$2.23^{+0.47}_{-0.35}\times10^{21}$ & 
$2.29^{+0.91}_{-0.91}\times10^{22}$ & 
$0.39_{-0.22}^{+0.19}$ & 
$1.83_{-0.24}^{+0.31}$ & 
$3.93^{+2.49}_{-1.22}\times10^{-4}$ & \dots & \dots & \dots & \dots & \dots\\
N4596\tablenotemark{a} & 2928 & 9.2 & \dots &
$2.70^{+5.94}_{-2.70}\times10^{21}$ & \dots & \dots &
$4.08_{-1.80}^{+4.38}$ & 
$5.28^{+19.72}_{-5.28}\times10^{-6}$ & \dots & \dots & \dots & \dots & \dots\\
N4649\tablenotemark{a} & 8182 & 52.4 & $30.8 / 37$ &
$1.54^{+1.25}_{-1.25}\times10^{21}$ & \dots & \dots &
$2.45_{-0.60}^{+0.69}$ & 
$1.11^{+0.82}_{-0.45}\times10^{-5}$ & \dots & \dots & \dots & \dots & \dots\\
N4697 & 784 & 41.4 & $3.2 / 2$ &
$<4.20\times10^{20}$ & \dots & \dots &
$1.81_{-0.27}^{+0.36}$ & 
$2.82^{+0.47}_{-0.17}\times10^{-6}$ & \dots & \dots & \dots & \dots & \dots\\
N4945 & 864 & 50.9 & $19.4 / 15$ &
$1.04^{+0.53}_{-0.35}\times10^{23}$ & 
$1.27^{+0.00}_{-1.27}\times10^{24}$ & 
$0.78_{-0.78}^{+0.22}$ & 
$0.48_{-0.88}^{+1.41}$ & 
$6.63^{+151.41}_{-5.97}\times10^{-5}$ & \dots & \dots & 
$6.20_{-0.01}^{+0.01}$ & 
$6.83^{+2.76}_{-2.16}\times10^{-2}$ & 
$5.64^{+26.35}_{-3.71}\times10^{-5}$\\
N5128 & 3965 & 49.5 & $255.8 / 200$ &
$5.96^{+0.59}_{-0.53}\times10^{22}$ & \dots & \dots &
$-0.05_{-0.11}^{+0.12}$ & 
$8.45^{+2.18}_{-1.59}\times10^{-5}$ & \dots & \dots & \dots & \dots & \dots\\
N5252 & 4054 & 60.1 & $676.7 / 445$ &
$0.00^{+0.00}_{-0.00}\times10^{19}$ & 
$1.94^{+0.06}_{-0.05}\times10^{22}$ & 
$0.97_{-0.00}^{+0.00}$ & 
$0.84_{-0.02}^{+0.03}$ & 
$7.52^{+0.36}_{-0.10}\times10^{-4}$ & \dots & \dots & \dots & \dots & \dots\\
N5845 & 4009 & 30.0 & $0.4 / 2$ &
$2.56^{+0.99}_{-1.48}\times10^{21}$ & \dots & \dots &
$2.52_{-0.76}^{+0.86}$ & 
$9.87^{+7.65}_{-4.12}\times10^{-6}$ & \dots & \dots & \dots & \dots & \dots\\
N6251 & 4130 & 45.4 & $456.4 / 358$ &
$9.73^{+0.57}_{-0.57}\times10^{20}$ & 
$2.22^{+0.66}_{-0.79}\times10^{24}$ & 
$0.90_{-0.21}^{+0.09}$ & 
$1.56_{-0.02}^{+0.02}$ & 
$9.68^{+10.35}_{-6.44}\times10^{-3}$ & \dots & \dots & \dots & \dots & \dots\\
N7052\tablenotemark{a} & 2931 & 9.6 & $16.2 / 5$ &
$2.68^{+0.00}_{-0.00}\times10^{21}$ & \dots & \dots &
$3.81_{-0.00}^{+0.00}$ & 
$5.41^{+0.00}_{-0.00}\times10^{-5}$ & \dots & \dots & \dots & \dots & \dots\\
N7457\tablenotemark{a} & 4697 & 9.0 & \dots &
$9.78^{+43.94}_{-9.78}\times10^{20}$ & \dots & \dots &
$2.97_{-1.13}^{+2.37}$ & 
$2.04^{+6.92}_{-2.04}\times10^{-6}$ & \dots & \dots & \dots & \dots & \dots\\
N7582 & 436 & 13.4 & $131.8 / 95$ &
$<1.03\times10^{20}$ & 
$1.40^{+0.09}_{-0.08}\times10^{23}$ & 
$0.95_{-0.01}^{+0.01}$ & 
$0.50_{-0.10}^{+0.05}$ & 
$4.14^{+0.38}_{-0.72}\times10^{-4}$ & \dots & \dots & \dots & \dots & \dots
    \enddata
    \label{t:xray}
\tablecomments{
    \ifthenelse{\value{emulateapj} = 1}
{}
{\parbox[t]{\textwidth}}
{Results from X-ray spectral analysis.  First column gives galaxy name.
The second column gives \emph{Chandra} observation identification
number.  The third column lists exposure time in units of ks.  Fourth
column lists $\chi^2 / \nu$ where $\nu$ is the number of degrees of
freedom.  If the fit used $C$-stat statistics instead of $\chi^2$
statistics, then the third column is left blank.  Best-fit parameters
with 1$\sigma$ errors for each.  A blank entry in a given column
indicates that the given component was not part of the spectral model
used.  Galaxies with superscript ``a'' were only able to constrain an
upper limit to the flux.  The model for Circinus also included a
pileup model.}
}
    \end{deluxetable*}
    \ifthenelse{\value{emulateapj} = 1}
{
\clearpage
\end{landscape}
}
{
\clearpage
}

\ifthenelse{\value{emulateapj} = 1}
{
\LongTables
}
{}
\begin{deluxetable*}{@{\extracolsep{\fill}}l@{\extracolsep{0pt}}crrr@{$\pm$}lrr@{$\pm$}lr}
\tablecaption{Black Hole Data}
\tablewidth{\textwidth}
\tablehead{
\colhead{Galaxy} & &
\colhead{AGN Class.} &
\colhead{$D / \mathrm{Mpc}$} & 
\multicolumn{2}{c}{$\log\left(\mbh\right)$} & 
\colhead{$\log\left(L_R\right)$} & 
\multicolumn{2}{c}{$\log\left(L_X\right)$} &
\colhead{Ref.}}
\startdata
Circinus & * & S2\tablenotemark{a} &  4.0 &  6.23 & 0.088 & 37.73 & 41.48 & 0.034 & 1,2 \\
IC~1459 & *  & S3 & 30.9 &  9.44 & 0.196 & 39.76 & 40.86 & 0.014 & 3,4 \\
IC~4296 & *  &  & 54.4 &  9.13 & 0.065 & 38.59 & 41.31 & 0.044 & 5,6 \\
Sgr~A* & *  &  &  0.008 &  6.61 & 0.064 & 32.48 & 33.33 & 0.068 & 7,8 \\
NGC~0221 &   &  &  0.9 &  6.49 & 0.088 &  \dots & 36.17 & 0.059 & 9 \\
NGC~0224 &   & S3\tablenotemark{a} &  0.8 &  8.17 & 0.161 & 32.14 & \multicolumn{2}{l}{$<36.00$} & 10,11 \\
NGC~0821 &   &  & 25.5 &  7.63 & 0.157 &  \dots & 38.44 & 0.640 & 12 \\
NGC~1023 &   &  & 12.1 &  7.66 & 0.044 &  \dots & 38.80 & 0.066 & 13 \\
NGC~1068 & *  & S2\tablenotemark{a} & 15.4 &  6.93 & 0.016 & 39.18 & 39.54 & 0.024 & 14,15 \\
NGC~1300 &   &  & 20.1 &  7.85 & 0.289 &  \dots & \multicolumn{2}{c}{\dots} & 16 \\
NGC~1399 &   &  & 21.1 &  8.71 & 0.060 &  \dots & \multicolumn{2}{l}{$<38.64$} & 17 \\
NGC~2748 &   &  & 24.9 &  7.67 & 0.497 &  \dots & \multicolumn{2}{c}{\dots} & 16 \\
NGC~2778 &   &  & 24.2 &  7.21 & 0.320 &  \dots & \multicolumn{2}{c}{\dots} & 12 \\
NGC~2787 & *  & S3b &  7.9 &  7.64 & 0.050 & 36.52 & 38.70 & 0.059 & 18,19 \\
NGC~3031 & *  & S1.8\tablenotemark{a} &  4.1 &  7.90 & 0.087 & 36.97 & 40.84 & 0.097 & 20,21 \\
NGC~3115 &   &  & 10.2 &  8.98 & 0.182 &  \dots & 38.04 & 0.312 & 22 \\
NGC~3227 & *  & S1.5 & 17.0 &  7.18 & 0.228 & 37.72 & 41.55 & 0.046 &  23,21 \\
NGC~3245 & *  & S3\tablenotemark{a} & 22.1 &  8.35 & 0.106 & 36.98 & 39.28 & 0.420 & 23,24 \\
NGC~3377 &   &  & 11.7 &  8.06 & 0.163 &  \dots & 38.00 & 0.322 & 12 \\
NGC~3379 & *  & S3\tablenotemark{a} & 11.7 &  8.09 & 0.250 & 35.81 & 38.17 & 0.205 & 25,26 \\
NGC~3384 &   &  & 11.7 &  7.25 & 0.042 &  \dots & \multicolumn{2}{l}{$<38.55$} & 12 \\
NGC~3585 &   &  & 21.2 &  8.53 & 0.122 &  \dots & 38.98 & 0.161 & 27 \\
NGC~3607 &   & S2 & 19.9 &  8.08 & 0.153 &  \dots & \multicolumn{2}{l}{$<38.60$} & 27 \\
NGC~3608 &   & S3\tablenotemark{a} & 23.0 &  8.32 & 0.173 &  \dots & \multicolumn{2}{l}{$<38.79$} & 12 \\
NGC~3998 & *  & S3b & 14.9 &  8.37 & 0.431 & 38.03 & 41.44 & 0.007 & 28,29 \\
NGC~4026 &   &  & 15.6 &  8.33 & 0.109 &  \dots & \multicolumn{2}{l}{$<38.53$} & 27 \\
NGC~4258 & *  & S2 &  7.2 &  7.58 & 0.001 & 36.03 & 40.83 & 0.096 & 30,21 \\
NGC~4261 &   & S3h & 33.4 &  8.74 & 0.090 & 39.32 & \multicolumn{2}{l}{$<40.92$} & 31,29 \\
NGC~4291 &   &  & 25.0 &  8.51 & 0.344 &  \dots & \multicolumn{2}{c}{\dots} & 12 \\
NGC~4342 &   &  & 18.0 &  8.56 & 0.185 &  \dots & 39.13 & 0.151 & 32 \\
NGC~4374 & *  & S2 & 17.0 &  9.18 & 0.231 & 38.77 & 39.42 & 1.503 & 33,34 \\
NGC~4459 &   & S3\tablenotemark{a} & 17.0 &  7.87 & 0.084 & 36.13 & \multicolumn{2}{l}{$<38.97$} & 18,24 \\
NGC~4473 &   &  & 17.0 &  8.11 & 0.348 &  \dots & \multicolumn{2}{l}{$<38.50$} & 12 \\
NGC~4486 & *  & S3 & 17.0 &  9.56 & 0.126 & 39.83 & 40.46 & 0.015 & 35,36 \\
NGC~4486A &   &  & 17.0 &  7.13 & 0.146 &  \dots & \multicolumn{2}{l}{$<38.96$} & 37 \\
NGC~4564 &   &  & 17.0 &  7.84 & 0.045 &  \dots & \multicolumn{2}{l}{$<38.79$} & 12 \\
NGC~4594 & *  & S1.9 & 10.3 &  8.76 & 0.413 & 37.89 & 40.19 & 0.307 & 38,39 \\
NGC~4596 &   & S3\tablenotemark{a} & 18.0 &  7.92 & 0.162 &  \dots & \multicolumn{2}{l}{$<38.72$} & 18 \\
NGC~4649 &   &  & 16.5 &  9.33 & 0.117 & 37.45 & \multicolumn{2}{l}{$<38.95$} & 12,40 \\
NGC~4697 &   &  & 12.4 &  8.29 & 0.038 &  \dots & 38.25 & 0.745 & 12 \\
NGC~5077 &   & S3b & 44.9 &  8.90 & 0.221 &  \dots & \multicolumn{2}{c}{\dots} & 12 \\
NGC~5128 & *  & S2? &  4.4 &  8.48 & 0.044 & 39.85 & 40.22 & 0.085 & 41,42 \\
NGC~5576 &   &  & 27.1 &  8.26 & 0.088 &  \dots & \multicolumn{2}{c}{\dots} & 27 \\
NGC~5845 &   &  & 28.7 &  8.46 & 0.223 &  \dots & 39.07 & 0.722 & 12 \\
NGC~6251 & *  & S2 & 106.0 &  8.78 & 0.151 & 41.01 & 42.50 & 0.207 & 43,44 \\
NGC~7052 &   &  & 70.9 &  8.60 & 0.223 & 39.43 & \multicolumn{2}{l}{$<40.69$} & 45,46 \\
NGC~7457 &   &  & 14.0 &  6.61 & 0.170 &  \dots & \multicolumn{2}{l}{$<38.28$} & 12 \\
NGC~7582 & *  & S2\tablenotemark{a} & 22.3 &  7.74 & 0.104 & 38.55 & 41.69 & 0.208 & 47,48 \\
PGC~49940 &   &  & 157.5 &  9.59 & 0.056 &  \dots & \multicolumn{2}{c}{\dots} & 5 \\
\hline &  \\
Cygnus~A &   & S1.9 & 257.1 &  6.43 & 0.126 & 41.54 & 44.23 & 0.088 & 49,6 \\
NGC~4151 &   & S1.5 & 13.9 &  7.65 & 0.048 & 38.20 & 41.69 & 0.074 & 50,21 \\
NGC~4303 &   & S2 & 17.9 &  6.65 & 0.349 & 38.46 & 38.89 & 0.124 & 51,52 \\
NGC~4742 &   &  & 16.4 &  7.18 & 0.151 &  \dots & \multicolumn{2}{c}{\dots} & 53 \\
NGC~4945 &   & S &  3.7 &  6.15 & 0.184 & 38.17 & 37.80 & 0.921 & 54,55 \\
NGC~5252 &   & S2 & 103.7 &  9.00 & 0.341 & 39.05 & 43.20 & 0.017 & 56,57 
\enddata
\label{t:data}
\tablecomments{This table lists all galaxies with dynamically measured
black hole masses.  Sources with an asterisk after their name are
those used in this paper.
The second column gives AGN classification from
\citet{2006A&A...455..773V} unless it has a superscript ``a,'' in
which case it comes from NED.  ``S1'' indicates type 1 (unobscured)
Seyfert, ``S2'' indicates type 2 (obscured) Seyfert, ``S1.X''
indicates transitional or intermediate Seyfert, ``S3'' indicates type
3 Seyfert or LINER galaxy, and ``?'' indicates that NGC~5128 is a
questionable BL~Lac object.  Note that NGC~3227 and NGC~4151 are
classified as type 1.5 but both have reverberation mapping masses
\citep{2003ApJ...585..121O,2006ApJ...651..775B} and thus have visible
broad-line regions.  Beyond these two galaxies, none of the galaxies
is obviously a Seyfert 1, though NGC~1068 and NGC~7582 are classified
as such by \citet{2006A&A...455..773V}.  We classify them according to
their NED classifications as Seyfert 2 galaxies.  NGC~1068 is a
Seyfert 2 galaxy and only shows broad Balmer lines in polarized light,
indicating that the light has been scattered and thus coming from
behind an obscured source \citep{1985ApJ...297..621A}.  NGC~7582 is a
classical Seyfert 2 galaxy that developed broad emission lines for a
short period of time in 1998 July \citep{1999ApJ...519L.123A}.  The
change to a Seyfert 1 spectrum may be explained by a stellar
disruption event, a change in the obscuring medium, or a type IIn
supernova \citep[][see also
\citealt{2000A&ARv..10...81V}]{1999ApJ...519L.123A}.  It has also been
suggested, based on its X-ray spectrum, that NGC~7582 is an obscured
narrow-line Seyfert 1 \citep{2005ApJ...625L..31D}.  For our purposes,
we classify this source as a Seyfert 2.  The paucity of true Seyfert 1
galaxies in our sample is not surprising since such bright central
engines would compromise dynamical black hole mass measurements.
The third column gives distance to the galaxy in units of $\Mpc$,
which is used to scale all data.
The fourth column lists logarithmic black hole mass per unit solar
mass as compiled by \citet{gultekinetal09b}.
The fifth column gives logarithmic radio luminosity in units of
$\ergs$.  The radio data come from the compilation of \citet{ho02}
with the following exceptions: IC~4296, NGC~5128, NGC~7582, NGC~4303,
and NGC~5252.
The sixth column gives logarithmic X-ray luminosity in units of
$\ergs$, which come from this work except for Sgr~A*
\citep{2001Natur.413...45B} and the upper limit on NGC~0224
\citep{2005ApJ...632.1042G}.  We leave the column blank if there are
no archival data available.
The final column lists original references for the \mbh\ measurement
and radio luminosity, if present.
The bottom portion of the table gives data for when the black hole
mass may be wrong.  
For this paper we only use data from galaxies that are in the top
portion and that have both radio and X-ray detections.  Blank entries
indicate that there were no archival data available and may be
followed up with more observations.}
\tablerefs{
(1) \citealt{2003ApJ...590..162G}, 
(2) \citealt{1983ApJ...268L..79T},
(3) \citealt{2002ApJ...578..787C}, 
(4) \citealt{1989MNRAS.240..591S},
(5) \citealt{2008arXiv0809.0766D},
(6) \citealt{1999ApJ...526...60S}, 
(7) \citealt{ghezetal08} and \citealt{2008arXiv0810.4674G}, 
(8) \citealt{1983A&A...122..143E},
(9) \citealt{2002MNRAS.335..517V}, 
(10) \citealt{benderetal05}, 
(11) \citealt{1992ApJ...390L...9C},
(12) \citealt{2003ApJ...583...92G}, 
(13) \citealt{2001ApJ...550...75B}, 
(14) \citealt{2003AandA...398..517L}, 
(15) \citealt{1984ApJ...278..544U},
(16) \citealt{2005MNRAS.359..504A}, 
(17) \citealt{2007ApJ...671.1321G}, 
(18) \citealt{2001ApJ...550...65S}, 
(19) \citealt{1980A&AS...40..295H},
(20) \citealt{2003AJ....125.1226D}, 
(21) \citealt{2001ApJS..133...77H},
(22) \citealt{1999MNRAS.303..495E}, 
(23) \citealt{barthetal01}, 
(24) \citealt{1991AJ....101..148W},
(25) \citealt{2000AJ....119.1157G}, 
(26) \citealt{1989ApJ...347..127F},
(27) \citealt{gultekinetal09a}, 
(28) \citealt{2006AandA...460..439D}, 
(29) \citealt{1984ApJ...287...41W},
(30) \citealt{2005ApJ...629..719H}, 
(31) \citealt{1996ApJ...470..444F}, 
(32) \citealt{1999ApJ...514..704C}, 
(33) \citealt{1998ApJ...492L.111B}, 
(34) \citealt{1977MmRAS..84...61J},
(35) \citealt{1997ApJ...489..579M}, 
(36) \citealt{1991AJ....101.1632B},
(37) \citealt{nowaketal07}, 
(38) \citealt{1988ApJ...335...40K}, 
(39) \citealt{1984A&A...134..207H},
(40) \citealt{1986Natur.321..753S},
(41) \citealt{2005AJ....130..406S}, 
(42) \citealt{1994ApJS...91..111W}, 
(43) \citealt{1999ApJ...515..583F}, 
(44) \citealt{1986ApJ...305..684J},
(45) \citealt{1998AJ....116.2220V}, 
(46) \citealt{1987A&A...183..203M},
(47) \citealt{2006AandA...460..449W}, 
(48) \citealt{1994ApJS...90..173G}, 
(49) \phantom{\;} \citealt{2003MNRAS.342..861T},
(50) \citealt{onkenetal07},
(51) \citealt{2007AandA...469..405P}, 
(52) \citealt{1991ApJS...75.1011G}, 
%
(53) listed as in preparation in \citealt{tremaineetal02} but never published,
(54) \citealt{1997ApJ...481L..23G},
(55) \citealt{1997MNRAS.284..830E},
(56) \citealt{2005A&A...431..465C},
(57) \citealt{1996ApJS..106..399P}. 
}
\end{deluxetable*}

\label{lastpage}
\end{document}